\documentclass[12pt]{emulateapj} 
\usepackage{natbib}
\usepackage{epsfig} 
\usepackage{times}
\def\be{\begin{eqnarray}}   \def\ee{\end{eqnarray}}
\def\ben{\begin{eqnarray*}} \def\een{\end{eqnarray*}} 
\def\sec#1{Section~\ref{sec:#1}}
\def\fig#1{Figure~\ref{fig:#1}} 
\def\tab#1{Table~\ref{tab:#1}}
\def\equ#1{equation~(\ref{equ:#1})}
\def\secp#1{Section~\ref{sec:#1}}
\def\figp#1{Figure~\ref{fig:#1}} 
\def\tabp#1{Table~\ref{tab:#1}}
\def\equp#1{equation~\ref{equ:#1}}

\newcommand{\Gyr}{\>{\rm Gyr}} 
\newcommand{\kpc}{\>{\rm kpc}}

\newcommand{\Msun}{\>{\rm M_{\odot}}} 
\newcommand{\Lsun}{\>{\rm L_{\odot}}} 
\newcommand{\degree}{\ensuremath{^\circ}}
\begin{document}
% \submitted{}
\title {Group Finding in the Stellar Halo using Photometric Surveys: Current Sensitivity and Future Prospects}

\author{Sanjib Sharma}
\affil{Department of Astronomy, Columbia University, New York, NY-10027
} 
\affil{Sydney Institute for Astronomy, School of Physics, University of Sydney, NSW 2006, Australia} 
\author{Kathryn V Johnston}
\affil{Department of Astronomy, Columbia University, New York, NY-10027
} 
\author{Steven R.  Majewski}
\affil{Department of Astronomy, University of
  Virginia,Charlottesville, VA- 22904}  
\author{James Bullock}
\affil{Center for
  Cosmology, Department of Physics \& Cosmology,University of California
  Irvine,CA-92697} 
\author{Ricardo R. Mu\~noz}
\affil{Department of Astronomy,
  Yale University,New Haven, Connecticut-06520}

% -------------------------------------------------------------------------
\begin{abstract}
  The Sloan Digital Sky Survey (SDSS) and the Two Micron All-Sky
  Survey (2MASS) provided the first deep and global photometric
  catalogues of stars in our halo and clearly mapped not only its
  structure but also demonstrated the ubiquity of substructure within
  it.  Future surveys promise to push such catalogues to ever
  increasing depths and larger numbers of stars.  This paper examines
  what can be learnt from current and future photometric databases
  using group-finding techniques.  We compare groups recovered from a
  sample of M-giants from 2MASS with those found in synthetic surveys
  of simulated $\Lambda$CDM stellar halos that were built entirely
  from satellite accretion events and demonstrate broad consistency 
  between the properties of the two sets. We also find that these 
  recovered groups are likely to represent the majority of 
  high-luminosity ($L>5 \times 10^6  L_{\odot}$) 
  satellites accreted within the last 10 Gyr and on orbits with
  apocenters within 100 kpc.  However the sensitivity of the M-giant
  survey to accretion events that were either ancient, from
  low-luminosity objects or those on radial orbits is limited because
  of the low number of stars, bias towards high metallicity stars and
  the shallow depth (distance explored only out to 100 kpc from the
  Sun).  We examine the extent to which these limitations are
  addressed by current and future surveys, in particular catalogues of
  main sequence turn off (MSTO) stars from SDSS and LSST, and of RR
  Lyrae stars from LSST or PanSTARRS.  The MSTO and RR Lyrae surveys are more
  sensitive to low luminosity events ($L \sim 10^5 L_{\odot}$ or less)
  than the 2MASS M-giant sample. Additionally, RR Lyrae surveys, with
  superior depth, are also good at detecting events on highly eccentric
  orbits whose debris tends to lie beyond 100 $\kpc$.  When combined
  we expect these photometric surveys to provide a comprehensive
  picture of the last 10 Gyr of Galactic accretion. Events older than
  this are too phase mixed to be discovered.  Pushing sensitivity back
  to earlier accretion times would require additional dimensions of
  information, such as velocity and metallicity or abundance
  measurements.
\end{abstract}

\section{Introduction} 
Over the last two decades, our view of the stellar halo has changed
from a diffuse, featureless cloud of stars surrounding the Galaxy, to
one crossed by many large-scale features such as the tidal tails of
the Sagittarius dwarf galaxy
\citep{1994Natur.370..194I,1995MNRAS.277..781I,2003ApJ...599.1082M},
the Virgo overdensity \citep{2008ApJ...673..864J} the
Triangulam-Andromeda structure
\citep{2004ApJ...615..732R,2004ApJ...615..738M,2007ApJ...668L.123M}
and the low latitude Monocerous ring
\citep{2002ApJ...569..245N,2003ApJ...588..824Y}.  The mapping of these
low surface brightness structures can be attributed to the advent of
large scale surveys such as the Two Micron All Sky Survey (2MASS) and
the Sloan Digital Sky Survey (SDSS) that have large numbers of stars
in their catalogues. Future surveys, such as GAIA \citep{perryman02},
LSST \citep{2009AAS...21346003I} SkyMapper \citep{2007PASA...24....1K}
and PanSTARRS, will map the stellar halo in ever more detail.

The presence of these structures lends support to the $\Lambda$CDM
model of galaxy formation in which the stellar halo is built up, at
least in part, hierarchically through mergers of smaller satellite
systems \citep{2001ApJ...548...33B}. However, simulating the stellar
halo in a cosmological context to test this picture is a challenging
task for two reasons.  First, the stellar halo is intrinsically faint,
containing only about $1$ per-cent of the total Milky Way
stars. Hence, to simulate the faint structures within the stellar halo
with adequate resolution requires enormous computation power. For
example, if a satellite with a stellar mass of $10^5 \Msun$ is
simulated with at least $10^3$ particles, then to simulate a whole
galaxy having stellar mass $10^{11} \Msun$ with the same mass
resolution will require a simulation with more than $10^9$ stellar
particles.  Second, the physics of
star formation and its feedback effects are complex phenomena that
have not yet been fully understood.

In recent years progress has been made in tackling both these issues.
Hydrodynamical simulations of galaxies including star formation and
feedback recipes are now being done \citep{2003ApJ...591..499A,
  2004ApJ...606...32R,2004ApJ...612..894B,2007MNRAS.374.1479G,
  2008MNRAS.389.1137S,2009ApJ...702.1058Z}. 
However, the highest resolution simulations
only have a stellar mass resolution of $10^4-10^5 \Msun$, which is not
enough to resolve the features in the stellar halo. Alternatively,
assuming that the stellar halo is built up entirely by means of
accretion events, hybrid techniques have been developed that use
collisionless simulations to follow the dynamical evolution of stars
in an analytical potential and a semi-analytical prescription to
incorporate star formation processes \citep{2005ApJ...635..931B}.
Although these hybrid techniques are not fully self-consistent they
can produce realistic models of the stellar halos with very high
resolution and can resolve even the lowest luminosity
structures. Recent improvements in hybrid techniques have implemented
the semi-analytical star formation recipes directly into cold dark
matter simulations \citep{2009MNRAS.397L..87L, 2008MNRAS.391...14D,2010MNRAS.406..744C}, so that with current dark matter simulations
like Via Lactea-II \citep{2008Natur.454..735D} and Aquarius
\citep{2008Natur.456...73S}, reaching a resolution of over $10^9$
dark matter particles within the virial radius, the future looks promising.

With the tremendous progress in both theory and observations it is now
possible to compare the two quantitatively. Since substructures in a
system are fluctuations in the density field one way to make this
comparison is to analyze the statistics of fluctuations. For example,
\cite{2008ApJ...680..295B} looked at the fractional root mean square
deviation of the stellar density from a smooth triaxial model in the
main sequence turn off stars (hereafter MSTO) of SDSS data. They
compared the radial dependence of these deviations with those apparent
in the $11$ simulated stellar halos of \cite{2005ApJ...635..931B} and
found the observations and simulations to be in rough agreement. One
could in principle also do a Fourier analysis of the fluctuations or
study the angular correlation function.  While such analyses are
useful for studying the global statistical properties of structures
they cannot say much about individual structures. Also when
calculating the rms deviations it is not clear if the deviations 
are due to structures or is it due to the inability of the analytical 
model to describe the smooth component of the halo.

An alternative approach to comparing fluctuations apparent in observed
and simulated stellar halos is to use group finding or clustering
algorithms.  The strength of the group-finding approach for this
particular problem is that, in addition to simple comparisons of group
properties, the results have a direct and simple physical
interpretation.  It was shown in \cite{2008ApJ...689..936J} that if
substructures standing out above the background density can be
identified (i.e. as groups) then the distribution of the properties of
these structures can in principle be related to a galaxy's accretion
history in terms of the characteristic epoch of accretion and the mass
and orbits of progenitor objects. Hence, group finding can be used to
constrain the accretion history of our Galaxy and compare it to
general expectations of the $\Lambda$CDM paradigm.

For example, in  \citet{2010ApJ...722..750S} we applied a 
hierarchical clustering
algorithm \citep{2009ApJ...703.1061S} to a sample of M-giants selected from the
2MASS catalogue and recovered sixteen structures, many of which were
previously known and associated with individual accretion events.  In
this paper, we go on to compare the properties of these recovered
structures with those found in synthetic surveys -- designed to mimic
the observations in the 2MASS catalogue -- of simulated stellar halos
\citep[taken from][]{2005ApJ...635..931B}, in order to assess how
similar they are.  We then ask what type of accretion events (in terms
of satellite accretion time, luminosity and orbit) are recovered in
the synthetic surveys, in order to assess how sensitive we expect the
2MASS survey be to our Galaxy's own history.

The group finder is also applied to synthetic surveys of the simulated
stellar halos chosen to mimic current and near-future photometric
catalogues of other stellar tracers.
For example, SDSS, which has a magnitude limit
of $r<22.5$, was able to map the stellar halo with MSTO stars out to
$36 \kpc$ with about 4 million MSTO stars
\citep{2008ApJ...680..295B}.
A similar MSTO generated from  LSST -- which will observe the sky in six
photometric bands, $ugrizy$, with a single visit limiting magnitude of
$r\sim24.5$ (or $r\sim27.5$ using co-added maps) -- will be capable of
probing the halo out to $100 \kpc$ with more than 20 million stars.
In addition, both LSST and PanSTARRS will
explore the transient optical sky and should be able to detect
variable RR Lyrae stars out to $400 \kpc$ and beyond, which is more
than the expected edge of the stellar halo.  

For group finding we use the code EnLink \citep{2009ApJ...703.1061S},
which is a density-based hierarchical group-finder. As noted in
\citet{2010ApJ...722..750S}, this code is ideally suited for 
this application for
four reasons.  First, structures in the stellar halo have arbitrary
shapes and a density-based group-finder is able to identify just such
irregular groups. Second, while many density-based group-finders
consider only groups formed above a fixed isodensity contour, EnLink's
clustering scheme can identify groups at all density levels.  This is
essential because stellar halo structures span a wide range of
densities that cannot be separated by a single isodensity contour.
Third, halo structures can have nested substructures and EnLink's
organizational scheme allows the detection of this full hierarchy of
structures.  Finally, the group-finder gives an estimate of the
significance of the groups, so spurious clusters can be ignored.

The paper is organized as follows. \sec{datasets} describes the data
sets used in the paper. \sec{methods} discusses the methods employed
for analyzing the data--- i.e., group finding.  In \sec{results1} we
ask how much any photometric survey can tell us about substructures in
the stellar halo by looking at an idealized survey, simulated
in the absence of observational errors. In \sec{results2} we look at groups
derived from simulated surveys constructed under more realistic
circumstances. In \sec{results3} we discuss the implications of these
results for groups recovered from current surveys of M-giant stars
(from 2MASS) and MSTO stars (from SDSS) in the real Universe. Finally
we summarize our findings in \sec{summary}.

% ---------------------------------------------------------------------------------
\section{Data} \label{sec:datasets}
\subsection{The 2MASS M-giant data}
We use a catalog of $59392$ M-giants identified from the 2MASS survey 
in a companion paper \citet{2010ApJ...722..750S} (Paper I).
A brief description of the catalogue and the selection procedure is
repeated below.
The 2MASS all sky point source catalog contains about $471$ million objects
(the majority of which are stars)
with precise astrometric positions on the sky and photometry in three bands
$J, H, K_s$. The survey catalog is $99\%$ complete for $K_s<14.3$.  
An initial sample of candidate M-giants was
generated by applying the selection criteria: 
\be
10.0 & <  & K_s < 14.0 \\
J-K_s & > & 0.97 \\
J-H   & < & 0.561(J-K_s)+0.36 \\
J-H & > & 0.561(J-K_s)+0.19  \\
K_s{\rm sin}(b) & > & 14.0 {\rm sin}(15^{\circ}).
\ee 
All magnitudes in the above equations are in
the intrinsic, dereddened 2MASS system (labeled with subscript 0
hereafter), with dereddening applied using the
\cite{1998ApJ...500..525S} extinction maps.  These selection criteria and the
dereddening method are similar to those used by \cite{2003ApJ...599.1082M} to
identify the tidal tails of Sagittarius dwarf galaxy. 

Distance estimates for these
stars were made by assuming a uniform metallicity of $[{\rm Fe/H}]=-1.0$ and an age of
$13.1 \Gyr$ and  calculating a linear fit to the color 
magnitude relation of the giant stars
using a theoretical isochrone from the Padova database  \citep{1994AAS..106..275B, 2008AA...482..883M,
  2004AA...415..571B}.  The best fit relationship is $M_K=3.26-9.42(J-K_s)_0$
where $M_K$ is the absolute magnitude of the star and $(J-K_s)_0$ its
color. Using this relation the distance modulus is then given by 
\be  
\mu=(K_s)_0-3.26+9.42(J-K_s)_0.
\label{equ:dmod_2mass}
\ee 
The specific values of age and metallicity were adopted because a) 
they roughly correspond to the expected values for the  
stellar halo  and b) they lead to the identification of the 
largest known stellar halo structure in the data, 
the tails of the Sagittarius dwarf galaxy,  with maximum 
clarity.

\subsection{Simulations}
\label{sec:sims}
To make theoretical predictions of structures in the stellar halo, 
we use the eleven stellar
halo models of \cite{2005ApJ...635..931B}, which were simulated within the
context of the $\Lambda$CDM cosmological paradigm.
These simulations follow the accretion of individual satellites modeled as
$N$-body particle systems onto a galaxy whose disk, bulge, and halo potential
is represented by time dependent analytical functions. Semi-analytical
prescriptions are used to assign a star formation history to each satellite
and a leaky accreting box chemical enrichment model is used to calculate the
metallicity as a function of age for the stellar populations
\citep{2005ApJ...632..872R,2006ApJ...638..585F}. The three main model
parameters of an accreting satellite are the time since accretion, $t_{\rm
  acc}$, its luminosity, $L_{\rm sat}$, and the circularity of its orbit,
defined as $\epsilon=J/J_{\rm circ}$ ($J$ being the angular momentum of the
orbit and $J_{\rm circ}$ the angular momentum of a circular orbit having same
energy). The distribution of these three parameters describes the accretion
history of a halo. To study the sensitivity of the properties of structures in
the stellar halo to accretion history, we additionally use a set of six
artificial stellar halo models (referred to as non-$\Lambda$CDM halos) from
\cite{2008ApJ...689..936J} that have accretion events that are predominantly
(i) {\it radial} ($\epsilon<0.2$), (ii) {\it circular} ($\epsilon>0.7$), (iii)
{\it old} ($t_{\rm acc}>11 \Gyr$), (iv) {\it young} ($t_{\rm acc}<8 \Gyr$), (v) {\it
  high-luminosity} ($L_{\rm sat}>10^7 \Lsun$), and (vi) {\it low-luminosity} ($L_{\rm sat}<10^7 \Lsun$).

\subsubsection{Generation of synthetic surveys from
  simulations}
\label{sec:sim2sur}
Each $N$-body particle in the simulations represents a population of stars
having a certain stellar mass, a distribution of ages and a monotonic
age-metallicity relation.  The real data from surveys on the other hand are in
the form of color-magnitude combinations of individual stars, shaped by
observational or other selection criteria. The procedure to convert a
simulated stellar halo model to a synthetic survey consists of four
steps \footnote{a code named {\sl Galaxia} was developed for
    this, details of which will be presented in a future paper}:
\begin{itemize}
\item {Spawning the stars}: We use the \cite{2001ApJ...554.1274C}
  exponential initial mass function (IMF) to generate stars corresponding to
  the given stellar mass of an $N$-body particle. If $M$ is the stellar mass
  associated with the $N$-body particle and $m_{\rm mean}$ the mean mass of
  the IMF distribution then the number of generated stars is given by
  approximately $M/m_{\rm mean}$. We use stochastic rounding to convert
  $M/m_{\rm mean}$ to an integer--- i.e., if the fractional part of $M/m_{\rm
    mean}$ is less than a
  Poisson distributed random number with a range between 0 and 1, we increment
  the integral part by 1. Each generated star is assigned a mass that is
  randomly drawn from the IMF, an age that is randomly drawn from the given
  age distribution and metallicity computed from the age-metallicity relation.
\item {Assigning color and magnitude}: Next, a finely spaced grid of
  isochrones obtained from the Padova database \citep{1994AAS..106..275B,
    2008AA...482..883M, 2004AA...415..571B,2004A&A...422..205G} is 
  interpolated to calculate the
  color and magnitude of each generated star. The absolute magnitude is
  converted to an apparent magnitude based on the distance of the $N$-body
  particle from a given point of observation. In this paper we assume the
  point of observation to be at $8.5 \kpc$ from the Galactic Center in the
  plane of the Galactic disk. Finally, depending upon the color magnitude
  limits of the survey being simulated, the stars are either accepted or
  rejected.
\item {Distributing the stars in space}: The position coordinates of first
  star that is spawned by an $N$-body particle, is assumed to be same as that
  of the spawning particle. Each subsequent star spawned by the $N$-body
  particle is distributed in a spherical region around the particle. The
  radius $R$ of the spherical region is determined by the $N$-body particle's
  $32$nd nearest neighbor in $({\rm X,Y,Z})$ position space. For this
  the multi-dimensional
  density estimation code EnBiD \citep{2006MNRAS.373.1293S} is used.  The
  stars are radially distributed according to the Epanechnikov kernel function
  $(1-r'^2)r'^2$ where $r'=r/R$ and $r'$ varies from $0$ to $1$. 
  A surface density map in X-Z plane of one of the  $\Lambda$CDM
  halos sampled using the above scheme is shown in \fig{f1}. It can
  be seen that the structures are quite adequately reproduced. 
\item {Observing our stars}: In simulations the distance of the stars is
  exactly known but in real surveys the distance is estimated by using
  observed properties of the selected stars. We use the same procedure to
  estimate the observed distances for the stars in our synthetic surveys.
\end{itemize}

\begin{table*}
  \centering
  \caption{\label{tab:tb1} Synthetic survey data sets generated from
    simulations.}
  \scriptsize
  \begin{tabular}{lcccccccccc}
    \hline
    Name & Type & Color limits & Apparent  & distance  & $f_{sample}$ &  $\sigma_{\mu}$ &Stars  & Groups & Purity \\ 
&    &  & Mag & limit $\kpc$&  & & & &  & 
    \\
    \hline
    S1      &Ideal    & $0.1<g-r<0.3$ & $r<27.5$   & 363  & 0.25 
    & 0.0    & $\sim 1.1 \times 10^7$  &62.4 &0.82  \\
    S2      &2MASS&$(J-K_s)_0>0.97$ &$(K_s)_0<14$ & 94.5 & 1.0 
    & 0.34   & $\sim 4.7 \times 10^4$  &8.4  &0.68  \\
    S3  &MSTO     & $0.1<g-r<0.3$ & $r<24.5$   & 100   & 0.25 
    & 0.64   & $\sim 9.1 \times 10^6$  &9.4  &0.69  \\
    S3$'$  &MSTO     & $0.1<g-r<0.2$ & $r<24.5$   & 100   & 1.0 
    & 0.58   & $\sim 5.9 \times 10^6$  &15.0 &0.66  \\
    S4      &RR Lyrae &               & $r<24.5$   & 600  &  
    & 0.11   & $\sim 6.2 \times 10^4$  &7.9  &0.69   \\
    S5  &SDSS(8000 sqd)  & $0.1<g-r<0.3$ & $r<22$   & 30   & 1.0  
    & 0.61   & $\sim 1.7 \times 10^6$  &2.3  &0.39  \\
   \hline
  \end{tabular}
\tablecomments{Each data set contains synthetic surveys of seventeen
  stellar halos,  of which eleven are with $\Lambda$CDM accretion
  history and six are with non-$\Lambda$CDM accretion history. The reported
  number of groups and purity are average over 11 $\Lambda$CDM halos.}
\end{table*}

\subsubsection{Synthetic surveys and data sets}
\label{sec:sur}
We use the simulations and the procedure described above to generate five
different types of synthetic surveys, namely S1, S2, S3, S4 and S5
(summarized in \tabp{tb1}). 
Each data set consists of surveys of
seventeen stellar halos, eleven of which correspond to halos having
$\Lambda$CDM accretion history ($\Lambda$CDM halos). The remaining 
six correspond to artificial accretion histories 
(non-$\Lambda$CDM halos see \secp{sims}) and
are named as radial, circular, old, young, high luminosity and low luminosity
halos to reflect the properties of the dominant accretion events in them.
Below we briefly describe these data sets and discuss their properties. 

\begin{itemize}
\item{\bf Data set S1 --- ``ideal'' surveys}

The first data set, S1, uses a color range of $0.1<g-r<0.3$ (color from SDSS
{\it ugriz} bands) to select MSTO stars and has an apparent magnitude limit of
$r<27.5$. The color and magnitude limits were chosen to encompass the
bulk of the stellar halo MSTO stars (since the main
sequence stars have a typical absolute magnitude of 4.7 this corresponds to a
distance limit of $\sim 363 \kpc$ for the survey). 
This data set is 
designed as an
idealized test case and hence we assume that the distances to the stars are
exactly known.  A comparison of the results of the group-finder applied to
this ideal data set with those from more realistic surveys allows us to
distinguish the limitations to sensitivity imposed by physical effects from
limitations that are due to observational constraints.  
For 
computational ease we sub-sample the data by a factor of 0.25, which
results in a sample size of about $10^7$. In Appendix \ref{sec:subsampling} we 
show that this sub-sampling has negligible impact on 
the analysis presented here.

\item{\bf Data set S2 --- 2MASS M-giant surveys}

The second data set, S2, mimics the 2MASS M-giant sample. 
Color  and 
magnitude limits of $(J-K_s)_0>0.97$ and  $(K_s)_0<14$ (2MASS $JHK$ band)
are used
to generate the sample. These limits are similar to those
used in \citet{2010ApJ...722..750S} to
identify structures in the stellar halo from 2MASS M-giants.
A magnitude limit of
$(K_s)_0<14$ means the data explores distances out to $\sim 95 \kpc$.

The distance modulus $\mu$ is calculated from the apparent magnitude 
of the star and its color using the relation 
\be
\mu=(K_s)_0-3.26+9.42(J-K_s)_0, 
\ee 
which corresponds to an isochrone of age 13.1 Gyr and metallicity
[Fe/H]=-1. See \citet{2010ApJ...722..750S} for further details.
However, this color-distance relationship is accurate only for a particular
metallicity. 
In real systems there is a range in the 
metallicities of the stars, so the  distance estimated using  
\equ{dmod_2mass} has some uncertainty associated with it. 
In order to evaluate the size of this error we calculated the dispersion 
 about
the actual value $\mu$ of stars separately for each satellite system  
in the eleven $\Lambda$CDM halos and then computed $\sigma_{\mu}$
as the mean value of this dispersion over all 
the satellites. 
(This approach avoids the computed value of $\sigma_{\mu}$ being dominated 
by the largest satellite system in the halo.) 
We found
$\sigma_\mu=0.34$ mag, in agreement with the estimates of 
\cite{2003ApJ...599.1082M} where they report $\sigma_{\mu}=0.36$  
for the 2MASS M-giants in the core of Sagittarius. 
This suggests that we can expect an uncertainty of 
about $18$ per-cent for the distance of M-giants.

\item{\bf Data sets S3 and S3' --- MSTO surveys}

The third data set S3 mimics an F and G-type MSTO star survey like the one possible with
LSST, assuming a magnitude limit of 24.5
\footnote{ This magnitude limit corresponds to the single visit depth of LSST:
  the co-added depth of LSST can reach up to 27.5} 
(which corresponds to a distance limit $\sim 100$ kpc) and 
a color range
$0.1<g-r<0.3$ chosen to match our ''ideal'' data set S1. 
For computational ease, like data set S1, data set S3 is also sub-sampled by a factor of 0.25 (see Appendix \ref{sec:subsampling} for further details). 
We generate a second set of MSTO surveys, S3', with a more
limited color range
 $0.1<g-r<0.2$ in order to 
explore how our results depend on our selection criteria. 
The photometric error was assumed to vary as 
\be
\Delta m_r=(0.04-0.039)x +0.039x^2 
\label{equ:merror}
\ee 
where $x=10.0^{0.4(m_r-m_5)}$ and $m_5=24.5$, in accordance with LSST 
specifications \citet{2009AAS...21346003I}.

The distances for these data sets are estimated from the $r$-band apparent
magnitude $m_r$ using the relation $\mu=m_r-4.7$ for the distance modulus.
We estimated $\sigma_\mu$ for S3 and S3' using the same approach 
as for the S2 data set, and these are tabulated 
\tab{tb1}. 
For individual satellite systems 
$\sigma_{\mu}$ was also found to have a dependence
on the mean metallicity of the systems -- for color range $0.1<g-r<0.4$, with metallicity [Fe/H] decreasing from -1 to -1.6, $\sigma_{\mu}$ 
was found to increase from 0.7 mag to 0.82 mag. For [Fe/H] below
$-1.6$,  
$\sigma_{\mu}$ was roughly constant at a value of 0.82 mag but there was 
a scatter of about 0.06 mag. Hence for [Fe/H]=-2 our estimated value of 
$\sigma_{\mu}=0.82 \pm 0.06$ is in agreement  with $\sigma_{\mu}=0.9$ of MSTO 
stars reported  by \citep{2008ApJ...680..295B} for 
globular cluster systems with color range $0.2<g-r<0.4$ and  
metallicity ${\rm [Fe/H]} \sim -2$. 

Note \cite{2008ApJ...680..295B} while analyzing the stellar halo 
using SDSS had used a color limit of $0.2<g-r<0.4$ 
to isolate the MSTO stars. However, the spread in absolute magnitude 
of MSTO stars is known to increase with increase in $g-r$ color.  
Hence to improve the distance estimation we here adopt a color limit 
of $0.1<g-r<0.3$ (data set S3), which is shifted by 0.1 mag bluewards 
as compared to that of \cite{2008ApJ...680..295B}. 
One can possibly extend the blue limit even more bluewards, 
but there are very few MSTO stars with $g-r<0.1$, moreover, 
one increases the chance of contamination from non MSTO stars.

Since we use a single absolute magnitude to estimate the
distance,  small features in the color versus absolute magnitude diagram 
can generate duplicate groups for some very luminous structures. 
One such feature in the $0.1<g-r<0.3$ color range is due to the  
horizontal branch stars. To avoid this 
we restrict our analysis to stars with absolute 
magnitude $M_r>1$.

\item{\bf Data set S4 --- RR Lyrae surveys}

The fourth data set, S4, mimics an LSST RR Lyrae survey and is
generated by identifying the 
horizontal branch stars that intersect the instability strip in the
luminosity temperature diagram. 
The blue edge of the instability strip is given by equation 
\begin{eqnarray}
\log T_{BE}& = & 3.999 -0.079\log(L/\Lsun) \nonumber \\
& & +0.056\log(M/\Msun)+0.06Y
\end{eqnarray}
from \citet{1987A&AS...68..119C} and  
we assume the helium fraction to be $Y=0.2409$ and the mean mass of RR
Lyrae to be $M=0.645 \Msun$, 
similar to the values adopted 
by \citet{2004ApJS..154..633C} for their theoretical studies. 
The width of the strip $\Delta\log(T_{eff})$ is assumed to be 0.09, 
which is slightly higher than the values adopted by 
\citet{2004ApJS..154..633C}. The range of luminosity was assumed to be 
$1.4<\log(L/\Lsun)<2.0$. To avoid contamination with main sequence
stars the mass was assumed to lie in a mass range of $0.5<\log(M/\Msun)<1.0$.
Next, to match the number
of stars in the survey to the expected number of RR Lyraes in the
Milky Way we set the sampling fraction to $f_{sample}=2$. This 
generated about $6\times 10^4$ stars with a density of about 
1.5 per square degree in accordance with the results of 
\citet{2005AJ....129.1096I} on SDSS DR1 data and  \citet{2004AJ....127.1158V}
on QUEST survey (about 1 to 1.3 per square degree).
The apparent magnitude limit was assumed to be $m_r=24.5$, which is
within the design limits of LSST and PanSTARRS.  
The RR Lyrae stars were found to have a mean absolute magnitude 
of about $M_r=0.55$, which implies a distance limit  
of $600 \kpc$ for the survey --- beyond the outer edge of he stellar halo. 

For our analysis we assume a distance uncertainty of 5 per-cent for the RR Lyraes
(i.e. scattering the true distances of stars in our surveys 
with a dispersion $\sigma_r/r=0.05$). 
This value may be slightly optimistic given the 
present state of the art measurements but probably within reach 
in future. If the period and
metallicity of an RR Lyrae can be 
accurately measured 
then its distance can be estimated to better than 5 per-cent
accuracy \citep{2008ApJS..179..242C}. 
However \cite{2009MNRAS.398.1757W} report a distance 
measurement of about 8 per-cent accuracy for RR Lyraes in SDSS
using a technique that utilizes
the light-curve shape itself to estimate metallicities.

\item{\bf Data set S5 --- SDSS survey}
The last data set mimics an MSTO catalog from the  
Sloan digital sky survey (SDSS). 
Like data set S3 it uses 
a color limit of $0.1<g-r<0.3$ to isolate the MSTO stars. 
The magnitude limit of the survey was set to $r<22.0$ and it covers 
an area of 8000 square degrees towards the north galactic cap. The photometric 
error was assumed to vary as described by \equ{merror} but with
$m_5=22.6$, which roughly 
reproduces the SDSS errors (Fig-7 in \citet{2008ApJ...673..864J}).
\end{itemize}

\section{Methods}\label{sec:methods}

\subsection{Group finding}
\label{sec:group_finder}
In this paper we use the hierarchical group-finder EnLink
that can cluster a set
of data points defined over an arbitrary space
\citep[described in detail in][]{2009ApJ...703.1061S}.
Paper I contains 
a complete discussion of the optimum parameters
for the group finder, as well as the choice of 
coordinate system for the data.
We include a brief outline of each of these below.

EnLink identifies the peaks in the density
distribution in the data-space as group centers. The region
around each peak, which is bounded by an isodensity contour corresponding to
the density at the valley, is associated with the group. 
The number of neighbors
employed for this density estimation
was chosen to be $k=30$: a smaller
value makes the results of the clustering algorithm sensitive to noise in the
data, while a larger value means that small structures go undetected.  

EnLink is built around a metric that can automatically adapt to a given data set, and hence can be applied to spaces of arbitrary numbers and types of dimension.
However, it was shown in Paper I
that the most effective way
to deal with photometric data sets where the angular position of stars 
is well known, but the distance can be very uncertain is to transform 
to a Cartesian coordinate system and use a simple Euclidean metric.
Prior to the transformation, a new distance estimate
$r'=5(\log(r/(10{\rm pc})))-\mu_0$ is adopted,  where $\mu_{0}$ is
a constant that determines the degree to
which the radial dimension is ignored or used.
The optimum value of $\mu_0$ was found to be 8. Note, the clustering 
results were not overly sensitive to the exact choice of $\mu_0$.  

A group finding scheme in general also generates false groups due to 
Poisson noise in the data. In order to make meaningful comparison  
between different data sets with different number of data points 
it is imperative that one uses a consistent and objective scheme to
screen out spurious groups.  In EnLink spurious groups that
can arise due to Poisson noise in the data are screened out by
looking at the  significance 
\be S=\frac{\ln(\rho_{\rm
    max})-\ln(\rho_{\rm min})}{\sigma_{\ln \rho}}.
\label{equ:significance}
\ee 
The contrast  between the density at the peak ($\rho_{\rm max}$) and valley
($\rho_{\rm min}$) where it overlaps with another group can be thought of as
the signal of a group and the noise in this signal is given by the variance
$\sigma_{\ln(\rho)}$ associated with the density estimator.	
An advantage of such a definition is that the resulting distribution
of significance of spurious groups is almost a Gaussian and hence 
the cumulative number of groups as function of significance can be
written as 
\be
G(>S)=\left(1-{\rm erf}\left(\frac{S f_1}{\sqrt{2}}\right)\right)\frac{f_2 N}{k}
\label{equ:GS}
\ee 
where $N$ is the number of data points and $k$ the number 
of neighbors used for density estimation and $f_1$ and $f_2$ are
constants for a given  $k$ and the dimensionality of data $d$.
We adopt a threshold $S=S_{\rm Th}$ for each of our synthetic surveys
(depending on the number $N$ of stars in the survey) that satisfies
the condition $G(>S_{\rm Th})=0.5$. This means that the expected
number of spurious groups in our analysis is 0.5. 
In EnLink for a chosen  $S_{\rm Th}$, all groups with 
$S < S_{\rm Th}$ are denied the status of a group
and are merged with their respective parent groups.

\begin{figure}
  \centering \includegraphics[width=0.45\textwidth]{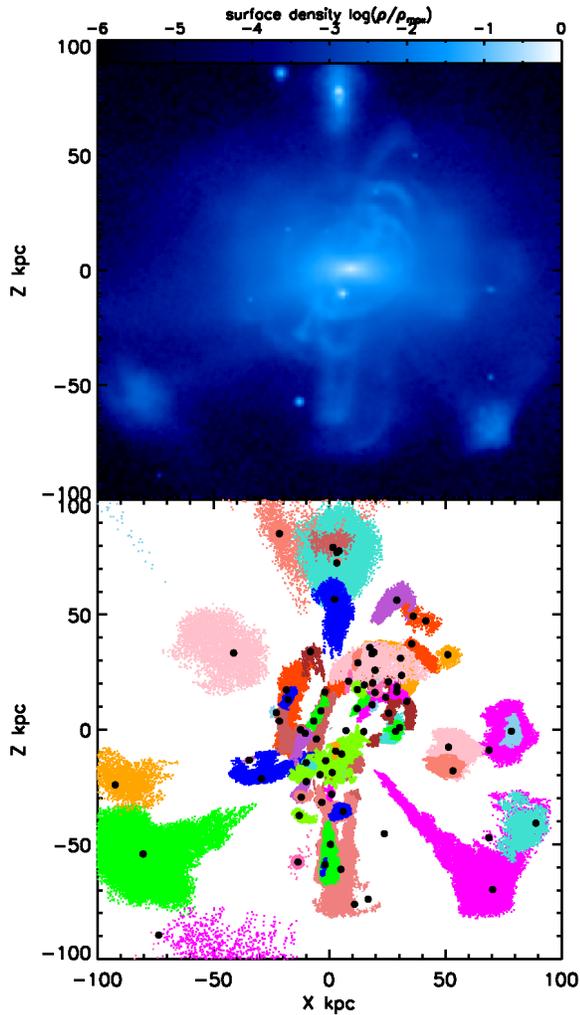}
\caption{ Plots showing the distribution of stars and groups in
  the X-Z space for one of the simulated stellar halo of data set S1. 
Top panel: surface density map of stars in the halo. Lower panel: 
a plot of stars which have been identified as groups by the group finder. 
Different colors in the plot represent different groups. The black filled
circles denote the point of peak density within a group. 
\label{fig:f1}}
\end{figure}

\begin{figure}
  \centering \includegraphics[width=0.45\textwidth]{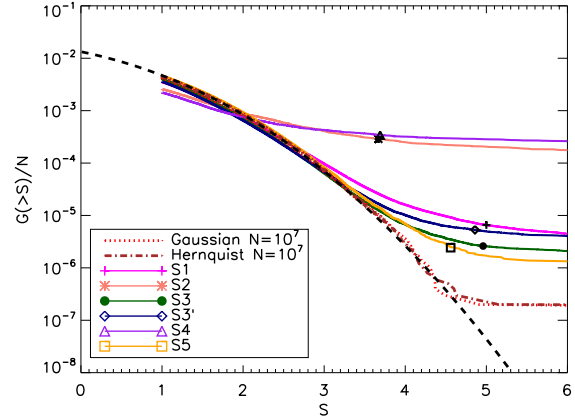}
\caption{ Cumulative distribution of significance of groups 
for  different data sets and its comparison to the expected
distribution for a structure-less halos (a Gaussian and a Hernquist
sphere). Group finding analysis is done in the modified radial
coordinates, except for the Gaussian sphere which is analyzed in the
normal co-ordinates. The adopted analytic relation is shown as the
dashed line. The symbols on the plot mark average (over 11 $\Lambda$CDM) 
significance threshold $S_{\rm Th}$ that is adopted for each data set. Only
groups with $S>1$ are shown in the plots.
\label{fig:f2}}
\end{figure}

A value of $f_1=1.0$ and $f_2=15.5/(d^{2.1}k^{0.2})$ 
were empirically derived by \cite{2009ApJ...703.1061S} and shown 
to be valid for a wide range of cases. An improved version
of the formula with $f_1=\sqrt{(d/4.0)(1-2.3/k)}$ and $f_2=0.4$ 
was presented in \cite{2010ApJ...721..582B}. 
But these formulas were derived for the case where the data is
generated by a homogeneous Poisson process and are reasonably accurate
for most applications. For non-homogeneous data the presence 
of density gradients on large scales makes the formula slightly 
inaccurate. Hence for greater precision one should derive the 
values of $f_1$ and $f_2$ for a model data whose large scale 
distribution is similar to the data being studied but 
otherwise does not have any substructures in it. We consider two 
models here a) a Gaussian sphere b) a Hernquist sphere  
with a scale radius of 15 kpc, which was shown to be an appropriate 
description of the stellar halo on large scales by \cite{2005ApJ...635..931B}. 
The sample size of models was set to $10^7$. 
A value of $f_1=0.93$ and $f_2=0.4$ was found to fit the distribution 
of significance of spurious groups and this is shown in \fig{f2}.
The distribution of significance averaged over $11$ $\Lambda$CDM halos
for the rest of the data sets is also shown. At small  $S$ 
the number of groups are dominated by the spurious groups  
and the curves lie on the predicted relationship. At large $S$ due to 
presence of real structures the curves flatten out. The symbols in the
figure denote the mean adopted value of $S_{Th}$ for each data set. 
The distribution of stars in data set S5 is very similar to data set 
S3 except for the fact that it is not over all sky.
Hence we revise the value of $S_{Th}$ for the data set S5 slightly 
from 4.55 as shown in the plot to 4.75. Note, for data sets with sample 
size less than $10^5$ the adopted relation is found to slightly 
underestimate the number of spurious groups but we ignore this.
Even if this fact is taken into account this would only lead to 
a very minor revision in values of $S_{\rm Th}$ for data sets S2 
and S4 and since the corresponding curves  are quite flat in this
region the number of detected groups would also not change much.

An example application of the group finder to one of the halos from 
data set S1 is shown in \fig{f1}. The upper panel shows the surface
density of stars in the X-Z plane while the lower panel shows the 
stars which have been identified as groups. It can be seen that our
scheme is very successful in detecting the structures.  Even faint 
structures that are not easily visible in the surface density maps 
are detected by the group finder.

EnLink is a hierarchical group finder and the reported groups 
obey a parent child relationship, with one master group being at the
top. In our case this master group represents the
smooth component of the halo and the rest of the groups are considered 
as substructures lying within it. Note, that since we analyze
magnitude limited surveys we expect two density peak even in a smooth 
halo: one at the galactic center and the other at the location of
the sun due to the presence of a large number of low luminosity
stars. Hence we ignore all groups whose density peaks lie within 
5 kpc of sun or the galactic center.

\subsection{Evaluation of clustering} \label{sec:cluseval}
When we apply EnLink to our synthetic data sets, because we know the
progenitor satellites from which the stars originally came from, we can make some quantitative assessment of how well the group finder has performed.
Following Paper I, we define {\it purity} as the
maximum fraction of stars in a group that came 
from a single satellite.
If $I$ is the set of all satellites and $J$ is the set of all groups, then 
\begin{eqnarray}
{\rm Purity}(j) & = & \frac{{\rm max}_{i \in I}\{n_{ij}\}}{n_{.j}},
\end{eqnarray}
where $n_{ij}$ is the number of stars from satellite $i$ in group $j$ and $n_{.j}=\sum_i n_{ij}$ is the total number of number of stars
in that group.
The mean value of purity $P=\sum {\rm Purity}(j)/|J|$
is a good indicator of to what extent the recovered groups correspond to real physical associations.

% ---------------------------------------------------------------------------
\section{Results I: surveys in an idealized Universe} \label{sec:results1} 
  
As a crucial step towards determining how groups recovered from a
photometric survey can be related to accretion events in the Galaxy's
history we first isolate the limitations imposed by the phase-mixing
of debris, rather than observational constraints.  In \sec{gpS1} we
characterize
how the peak density of debris --- which determines whether it might
be detectable, whether by a group-finder as a group or by some other
means --- depends on the physical characteristics of an accretion
event.
We then apply what we have learned to understand which progenitor
satellites produce detectable groups in a full stellar halo
(\secp{gpS2}) and finally to outline the prospects for interpreting the
full distribution of group characteristics in terms of accretion
histories (\secp{prospects}). For the latter two sections we use the
idealized data set S1 (see \sec{sim2sur}), consisting of synthetic
surveys with MSTO stars as tracers but the distance
to the stars being assumed to be known exactly.

Note that we restrict our attention to the properties of groups in
stellar halos that correspond to unbound structures because the number
and distribution of the structures that are still bound (i.e., the
missing satellite problem of the $\Lambda$CDM paradigm) is still an
issue of much debate. In the simulations, the star formation recipes
were tuned to match the observed distribution of the still surviving
satellite galaxies in the Milky Way known at the time of the
publication of \cite{2005ApJ...635..931B}. Hence the number of
satellites are not a prediction of the models but rather a boundary
condition. The true prediction of the simulated models is the number
of unbound structures corresponding to the ancestral siblings of the
present day classical dwarf spheroidal satellites that are formed by
the dynamical interaction of these satellites with the potential of
the host galaxy.

\begin{figure}
  \centering \includegraphics[width=0.45\textwidth]{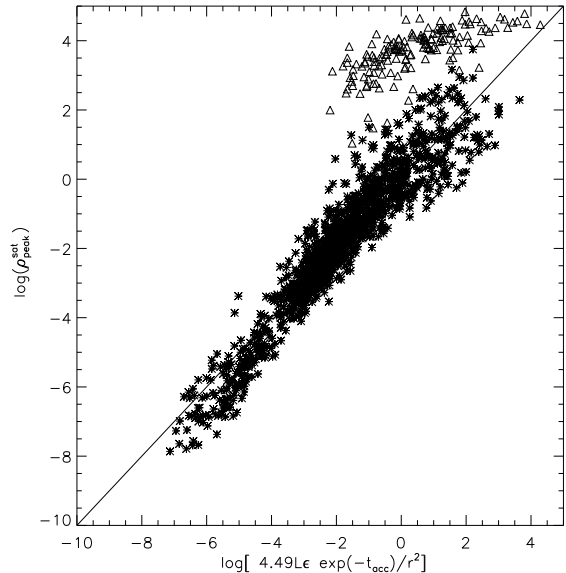}
\caption{ $\rho_{\rm peak}$ in arbitrary units for all 1515 simulated accretion events plotted as a function of 
$4.49e^{-t}\frac{L \epsilon}{r^2}$. The stars are for unbound structures
while the triangles are for bound structures. 
\label{fig:f3}}
\end{figure}

\subsection{Evolution of the peak density of satellite debris}
\label{sec:gpS1}
The peak density $\rho_{\rm peak}$ of an unbound structure evolves in
time from an initial maximum state set by the structural properties of
the satellite before disruption to a final minimum state where it is
fully mixed in the entire volume roughly outlined by the orbit of the
satellite. In the intermediate state, transient structures such as
streams, shells and clouds are apparent.  Below, we individually
analyze each simulated satellite debris (a total of 1515 satellites'
debris from our eleven stellar halo models) in order to understand the
dependence of peak density on the properties of its progenitor
satellite during each of these stages.

Initially the structure is bound and its peak density $\rho_{\rm
  peak}$ is observed to depend primarily upon the luminosity $L$ of
the progenitor satellite.  The exact dependence assumed in the
simulations is given by $\rho_{\rm peak} \propto L^{0.4}$ and comes
from fitting King profiles to observed dwarf galaxy data
\citep{1998ARAA..36..435M}. In the final, completely phase mixed state
the ensemble of stars in the original satellite occupy a large volume
$V_{mix}$ and have peak density of order $L/V_{mix}$. In a spherical
potential, $V_{mix}$ is expected to be inversely dependent on the
circularity of the orbit $\epsilon$ because the debris occupies the
region between the apocenter and the pericenter of the orbit and an
eccentric orbit (with lower $\epsilon$) explores a wider range in
radius. In the case of an axisymmetric potential the debris also
occupies the region generated by the precession of the orbit about the
axis of the system.  Evolution in the intermediate state is governed
by phase mixing, which occurs on a characteristic time scale $T_{mix}$
that depends on both the circularity $\epsilon$ and the mass of the
satellite \citep{1998ApJ...495..297J,1999MNRAS.307..495H}, which in
turn is correlated to luminosity $L$.  In this stage, the density has
both spatial and temporal dependence. Using the action angle formalism
\cite{1999MNRAS.307..495H} analytically deduced the general dependence
as $\rho_{\rm peak} \propto 1/(r^2 t_{\rm acc}^2)$ for spherical
potentials and as $\rho_{\rm peak} \propto 1/(r^2 t_{\rm acc}^3)$ for
axis-symmetric potentials, where $r$ is the distance of the peak
density of a satellite debris from the center of the parent galaxy and
$t_{\rm acc}$ is the time since accretion of the progenitor satellite.

Rather than model these multiple stages and effects independently, we
instead examine empirically to what extent the peak densities of
debris in our simulations can be connected to progenitor properties
with a single formula.  Overall, the physical effects described above
lead us to expect that $\rho_{\rm peak}$ follows a $1/r^2$ radial
dependence and is largest for objects that are luminous, recently
accreted and are disrupting on mildly eccentric orbits.  We find that
the relation \be \rho^{\rm emp}_{\rm peak} & \propto & e^{-t_{\rm
    acc}}\frac{L \epsilon}{r^2}
\label{equ:eq1}
\ee provides the least scatter about the actual value $\rho_{\rm
  peak}$ measured for the simulated satellite debris (\figp{f3}) and
is a reasonable representation of trends discussed earlier.  Note that
in \fig{f3}, instead of associating $\rho_{\rm peak}$ simply with the
global density maximum, we find the point where $\rho r^2$ is a
maximum, a choice that effectively identifies the point where the
density contrast with respect to the background is greatest.

\subsection{Probability of detecting accretion events as
groups}
\label{sec:gpS2}

\begin{figure*}
  \centering \includegraphics[width=0.9\textwidth]{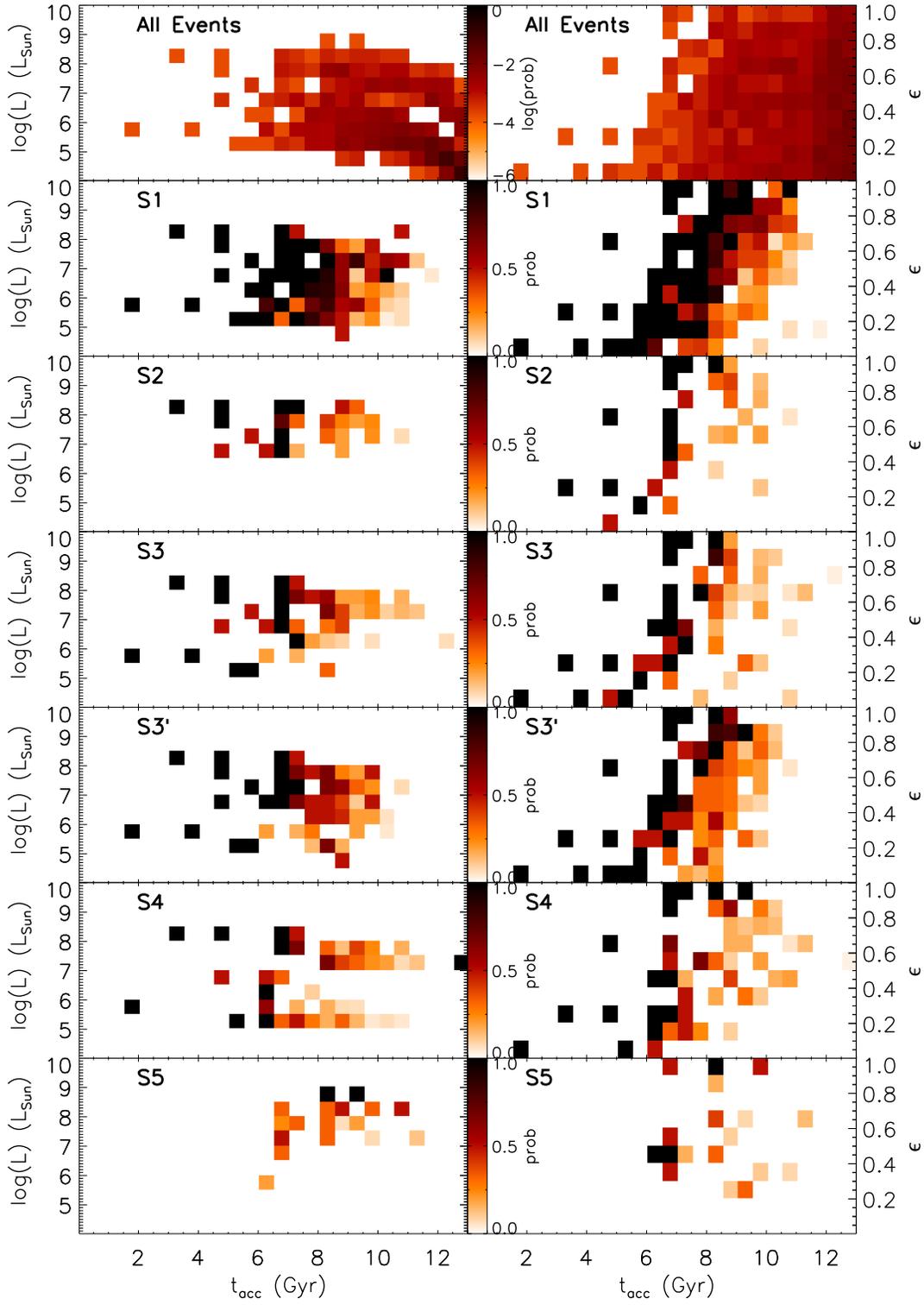}
\caption{ 
Probability distribution of detecting accretion events in the
$(L,t_{\rm acc})$ and $(\epsilon,t_{\rm acc})$ parameter space 
for various photometric surveys. Only events giving rise to unbound structures are shown.   
The type of data set is labeled on each panel, further details on labels
can be found in \tab{tb1}. The color scheme is such that the darkness 
increases monotonically with probability. The panels labeled all
events shows the probability
distribution of all the accretion events and is plotted in logarithmic
units (see color bar). 
The other columns show the conditional probability
of detecting the events in various data sets given the probability of all the
events and it is plotted in linear units (see color bar).
\label{fig:f4}}
\end{figure*}

The top panels of \fig{f4} shows the average accretion history adopted
in eleven $\Lambda$CDM model halos by plotting the probability of
occurrence of accretion events $P(E)$ (i.e. the number fraction of
events of a given $(L,t_{\rm acc}$ in the left hand panel, and of a
given $(\epsilon,t_{\rm acc})$ in the right hand panel).  Note the
distribution of $\epsilon$ is non-uniform for recent events because we
only included those that gave rise to unbound structures --- events on
circular orbits ($\epsilon \sim 1$) remain bound for a longer time
than those on more eccentric orbits ($\epsilon \sim 0$).  For $t<8
\Gyr$ the probability also abruptly goes to zero for $L< 10^5
\Lsun$. This is because recent accretion events with $L<10^5 \Lsun$
are not included in the simulations.

We can use our understanding of the evolution of the peak density of
satellite debris to develop some intuition for which of the events in
the top panels of \fig{f4} are most likely to be recovered as groups.
An unbound structure in a stellar halo is expected to be detected as a
group if the ratio $\rho_{\rm peak}/\rho_{b}>1$ (where $\rho_b$ is the
density of the background stars, i.e., halo stars 
after excluding the stream under consideration). 
Using \equp{significance} this
corresponds to a significance level of $S \sim 3$.  
The radial distribution of halo stars was 
explored in \citet{2005ApJ...635..931B} and they found that 
the  slope, $d\ln(\rho)/d\ln r$, transitions between -1 for $r<10$
kpc  to $-3.5$ for $r>50$ kpc . As a first order approximation we assume 
$\rho_{b} \propto 1/r^2$, which when combined with \equ{eq1}
gives for the ratio of peak to background density as 
\be 
\rho_{\rm peak}/\rho_{b} & \propto & e^{-t_{\rm acc}} L \epsilon
\label{equ:eq2}
\ee 
In the above equation, among the three parameters of an
accretion event, the strongest dependence is on
$t_{\rm acc}$; so the oldest accretion events have the least probability of
being detected. For given $t_{\rm acc}$, events with high luminosity and on
more circular orbits (high $\epsilon$) are more likely to be detected. 

We confirmed this intuition by applying the group finder to the
surveys if the eleven $\Lambda$CDM models our idealized data set S1
--- the second row of panels in in \fig{f4} shows the fraction of
accretion events in the $(L,t_{\rm acc})$ and $(\epsilon,t_{\rm acc})$
plane that were recovered as groups from these surveys.  Formally, if
$P(E)$ is the probability of occurrence of all the accretion events
and $P(SE)$ the probability of an event being identified by the
group-finder in a stellar halo, then the panels plot the probability
of detecting an event in a stellar halo, given by the conditional
probability $P(S|E)=P(SE)/P(E)$.  A comparison of the second and top
rows in the figure demonstrates that nearly all events on all types of
orbits and of all luminosities are recovered for $t_{\rm acc}< 8
\Gyr$.  Older events on more circular orbits are detectable as groups
even further back, as might be anticipated from \equ{eq2}. 
  However, few accretion events older than 10 Gyr ago were recovered
  and none older than 11 Gyr. We conclude that the phase-mixing of
  debris imposes this fundamental limit, and photometric surveys alone
  will never be able to explore this epoch of our Galaxy's accretion
  history.

\subsection{Relating group properties to accretion history.} \label{sec:prospects}

Having confirmed how the recovery of a disrupted satellite as a group
depends on its accretion characteristics, we now examine to what
extent the {\it distribution} of group properties in a stellar halo
reflects the {\it distribution} of properties of accreted objects, and
hence what we might be able to say about the recent Galactic accretion
history.

For example, simple intuition suggests that the number of groups
should reflect the number of recent events.  The top row of \tab{tb2}
shows that we recovered an average of 62 groups from our S1
$\Lambda$CDM data sets with a minimum of 36 and a maximum of 82.  As
anticipated, we recovered significantly more groups from those of our
non-$\Lambda$CDM that had a larger number of recent accretion events
(i.e. the ones biased towards recent accretion, or dominated by many
low-luminosity events) and fewer groups from those that had a smaller
number (i.e. the ones biased towards ancient accretion, or dominated
by few high-luminosity events).  The non-$\Lambda$CDM stellar halo
formed from predominantly circular events also gave rise to more
events as these can be detected for longer (as demonstrated in \sec{gpS2} 
and by \equ{eq2})

Similarly, since accretion time is the dominant factor that determines
the detectability of an accretion event, we expect the fraction of
halo material detected as groups to be primarily related to the
integrated mass accretion history of the halo and less sensitive to
the total number of such accretion events.  The top row of \tab{tb3}
shows that the fraction of stars in groups varies widely for our S1
$\Lambda$CDM data sets with a minimum of less than 2\% and a maximum
over 20\%: the total fraction is heavily influenced by the most
luminous object accreted on to the halo, and these are very few in
number.  The recently accreted halo (i.e., the young halo) shows the
maximum fraction in groups as the structures have not yet
dispersed. In spite of the total number of accretion events being very
different, the low luminosity and high luminosity halos have similar
fractions because for both of them a similar amount of mass was
accreted within the last $10 \Gyr$.  The circular halos, which can
probe much older events, have higher fractions compared to the radial
halos.

\begin{figure}
  \centering \includegraphics[width=0.49\textwidth]{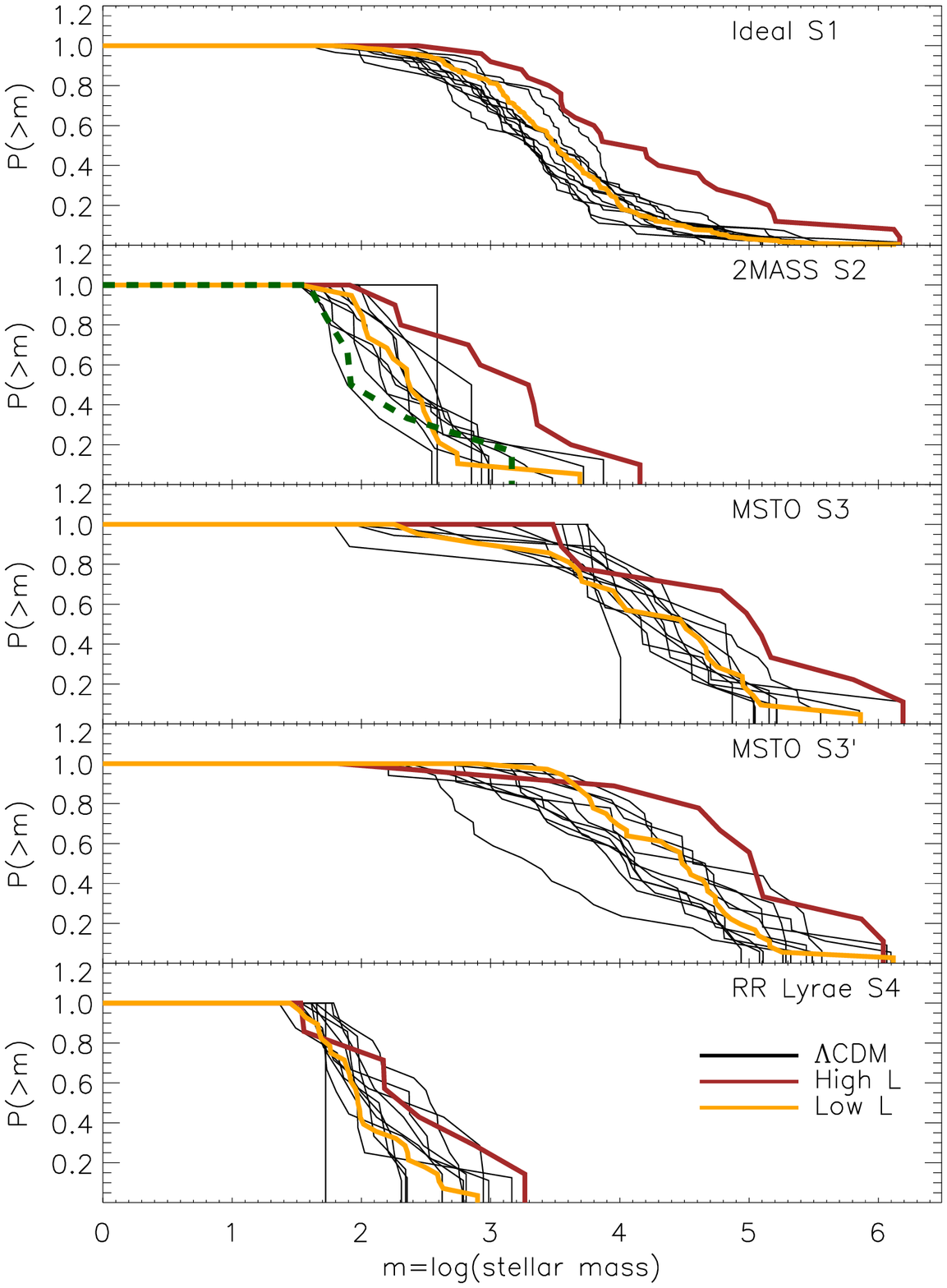}
\caption{Normalized cumulative distribution of stellar mass (number of
  stars in a group) in groups discovered
by the group-finder  for various photometric surveys.
 The curve corresponding to high
luminosity halo is easily distinguishable from other halos.
\label{fig:f5}}
\end{figure}

\begin{figure}
  \centering \includegraphics[width=0.49\textwidth]{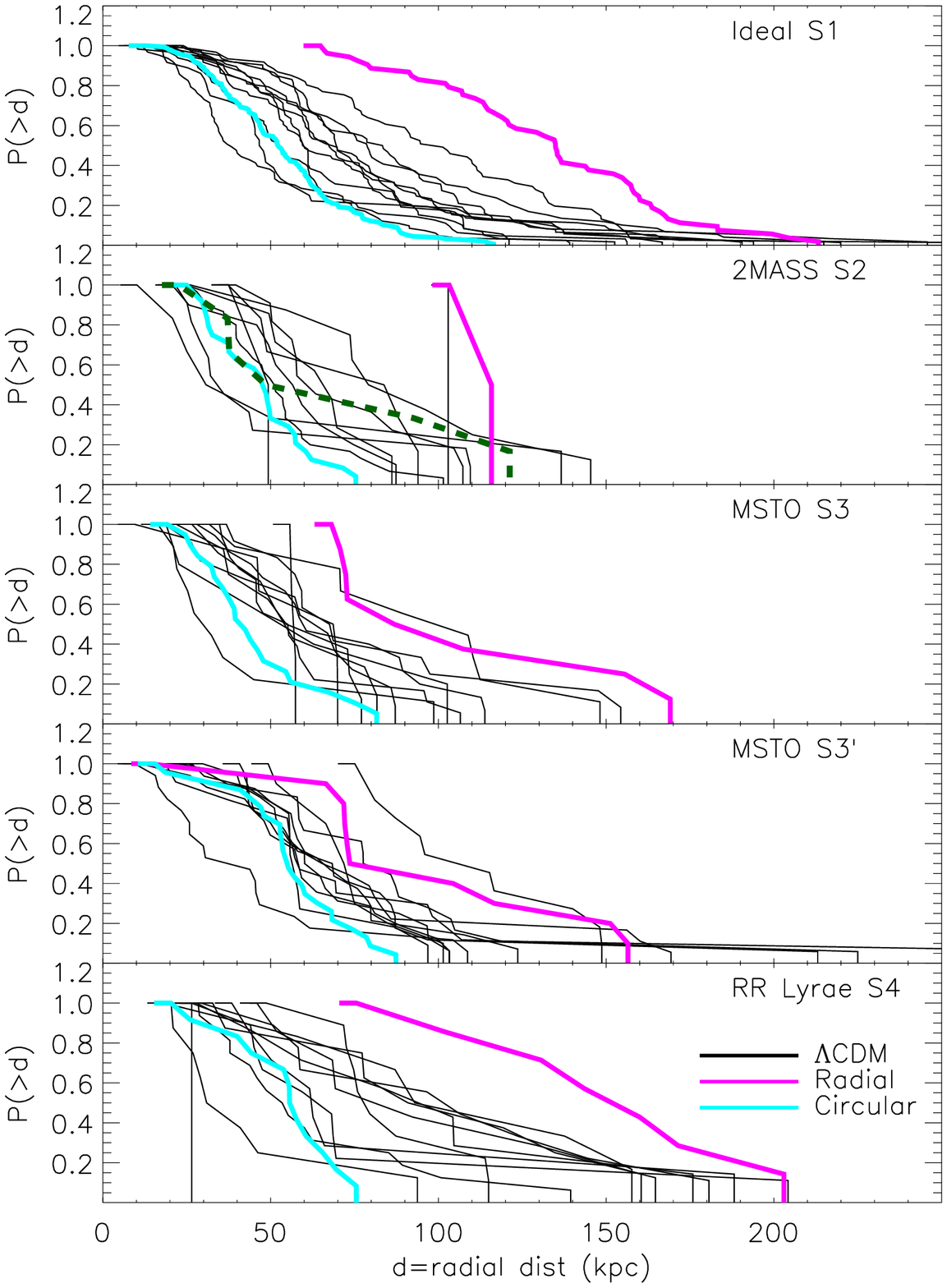}
\caption{Normalized cumulative distribution of radial distance (with respect to
the Sun) of  groups discovered by the group-finder  for various photometric surveys.
The distance of a group is defined as the heliocentric
distance to the point within a group that has maximum number density of stars.
The distributions of radial and circular halos are easily
distinguishable in the plots. Note, for the simulated halos the 
plotted distance is the actual distance and not the one inferred from 
properties of stars.
\label{fig:f6}}
\end{figure}

We would also like to know the luminosity function of accreted
objects, which should be reflected in the distribution of group
stellar masses $m_{\rm group}$.  The thin black lines in the top left
panel of \fig{f5} show the normalized, cumulative distribution of
group masses recovered from our 11 $\Lambda$CDM --- we found that our
non-$\Lambda$CDM halos that were constructed from events that had a
$\Lambda$CDM luminosity function (i.e. the old/recent and
circular/radial halos) all lay within the region occupied by the
-$\Lambda$CDM lines (i.e. they had the same {\it group mass
  function}).  In contrast, the brown line in this panel shows the
group mass function for the high-luminosity non-$\Lambda$CDM halo,
which is clearly distinct from the $\Lambda$CDM lines and biased
towards higher mass groups.  While the corresponding results for the
low-luminosity halo are less striking in this panel (yellow line),
\tab{tb2} already demonstrates that the difference in
luminosity function could be seen in the far larger
number of groups recovered from this halo than for the
$\Lambda$CDM halos.

The last unknown in the accretion history is the orbit distribution.
Since eccentric orbits with low $\epsilon$ have larger apocenters than
more circular orbits of the same energy and a larger fraction of time
during an orbit is spent near the apocenter rather than the
pericenter, groups formed from radial accretion events are expected to
be detected further away from the Galactic center. The probability distribution
of radial distance of groups for data set S1 shown in top left panel
of \fig{f6} demonstrates this.  The black lines are again
for the $\Lambda$CDM halos. The radial (purple) and the circular (cyan) halo
profiles clearly outline the black lines.
More than 2/3 of the groups in the radial halo are above $100
\kpc$ while 2/3 of the circular halo groups are below $60 \kpc$. For
the $\Lambda$CDM halos about $30\%$ of the groups lie above $100
\kpc$.

In summary, our analysis here using data set S1 suggests that, if the
distance of the stars is known accurately, then number of groups,
amount of material in substructures and the probability distributions
of masses and radial distance can be related to the number, luminosity
function and orbit distribution of recent accretion events.  The
remaining question is whether the accretion history of a halo can be
constrained with real observational data.

\begin{table}
  \centering
  \caption{\label{tab:tb2} Number of recovered groups for various
    surveys}
  \scriptsize
  \begin{tabular}{l|cccccc|lcc}
    \hline
   Sur   & Rad & Circ & Old  & You &  Hi & Lo & & $\Lambda$CDM & \\ 
   vey  &  &  &   & ng &  &   &  Mean & Min & Max \\ 
   \hline
S1  & 53 & 156 & 0 & 148 & 25 & 215 & 62.4  $\ \ \pm$ 12.3 & 36 & 82 \\
S2  & 2  & 24 & 0 & 33 & 10  & 19  & 8.4 $\ \pm$ 8.1 &  0 & 30 \\
S3  & 8  & 19 & 0 & 29 & 9  & 21  & 9.4  $\ \pm$ 4.4 &  3 & 18 \\
S3' & 10  & 23 & 2 & 33 & 9  & 36  & 15.0 $\pm$ 3.0 &  10 & 18 \\
S4  & 7  & 12 & 0 & 35 & 7  & 28  & 7.9  $\ \pm$ 3.5 &  1 & 16 \\
S5  & 1  & 3 & 0 & 8 & 0  & 3  & 2.3  $\ \pm$ 1.5 &  0 & 5 \\
   \hline
2MASS & & & &  & & & 6 & & \\
   \hline
  \end{tabular}
\end{table}

\begin{table*}
  \centering
  \caption{\label{tab:tb3} Fraction of material in recovered groups for various
    surveys}
  \scriptsize
  \begin{tabular}{l|cccccc|rccc|rccc}
    \hline
    & \multicolumn{10}{|c}{Unbound groups only} & 
%\multicolumn{4}{|c}{} &
    \multicolumn{4}{|c}{All groups, both bound and unbound} \\
    \hline
    Survey  & Rad & Circ & Old  & Young &  Hi & Lo & & $\Lambda$CDM &
    & & & $\Lambda$CDM  \\ 
       &  &  &   &  &  &   &  Mean  & Min & Max & stddev & Mean  & Min & Max & stddev \\ 
      &  &  &   &  &  &   &    & & & dex &    & & & dex \\ 
   \hline
S1  & 0.05 & 0.14 & 0.0 & 0.50 & 0.36 & 0.24 & 0.09  & 1e-2 & 0.23 & 0.34
&    0.36 &   0.10 &    0.53 &    0.26 \\
S2  & 8e-3 & 0.13 & 0 & 0.51 & 0.23 & 0.27  &  0.09  & 0.0 & 0.27 &0.76 
&    0.23 &   0.01 &    0.44 &    0.55 \\ 
S3 & 9e-3 & 0.04 & 0 & 0.38& 0.26 & 0.10  & 0.07 & 2e-3  & 0.20  & 0.59
&    0.19 &   0.03 &    0.38 &    0.37 \\
S3'  & 0.05 & 0.18 & 0.02 & 0.53 & 0.41 & 0.21 & 0.16  & 4e-2 & 0.42 & 0.31 
&    0.36 &    0.15 &    0.53 &    0.19 \\ 
S4  & 4e-3 & 0.02 & 0.0 & 0.48 & 0.05 & 0.04 & 0.03 & 9e-4  & 0.08 & 0.47
&   0.10 &   0.03 &    0.33 &    0.30 \\
S5  & 3e-5 & 0.02 & 0 & 0.10 & 0 & 1e-2 & 0.04 & 0  & 0.13 & 0.81
&   0.06 & 2e-4 &    0.13 &    0.97 \\
   \hline
2MASS &  & &  & &  & & 0.045 & & & &  & &  & \\
   \hline
  \end{tabular}
\end{table*}

\begin{figure}
  \centering \includegraphics[width=0.49\textwidth]{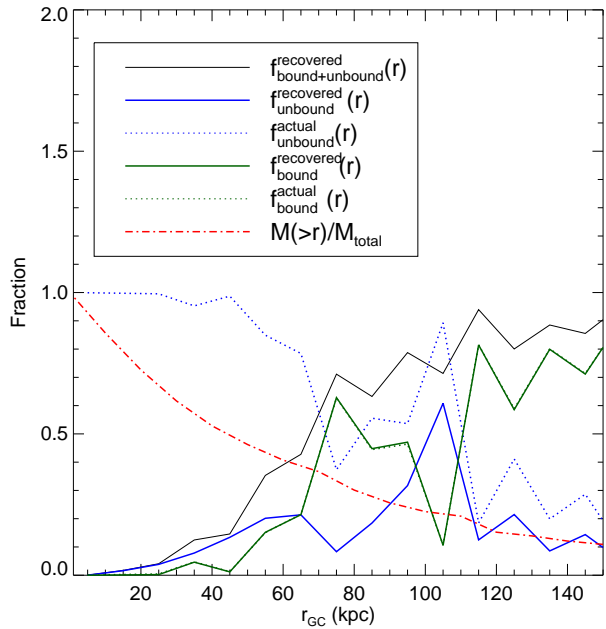}
\caption{ The fraction of material recovered as groups as a function of
  galactocentric radius for the idealized survey S1 (averaged over 
  11 $\Lambda$CDM halos).
  The fraction of material for bound and unbound
  groups is shown separately. The dotted curves show the actual fraction of
  material in bound and unbound structures in the simulations. 
  Note, the recovered and
  actual fraction of material in bound structures are indistinguishable in the
  plot. 
\label{fig:fraction}}
\end{figure}

\subsection{Fraction of material in groups}
Our results (top row \tab{tb3}) suggest that about 9\% of material is in
unbound recovered groups. As mentioned earlier fraction of material 
in groups is found to vary greatly from halo to halo and this is 
because it is typically  dominated by a few luminous accretion events.
A significant fraction of material is also
contained in bound groups. When these are taken into account the
fraction is found to rise to 36\% (last four columns of top row
in \tab{tb3}). Nevertheless, the fraction of material in unbound 
groups is still quite small. To explore 
this further, we plot  in \fig{fraction} the fraction of
material in recovered groups as a function of galactocentric 
radius (black solid line). The contributions of bound and unbound 
groups are also shown separately (blue and dark green solid lines). 
To create the plots the data points were tagged as recovered or
un-recovered and bound or unbound and then binned by radial 
galactocentric distances.
The dashed lines show the actual 
fraction of material in bound and unbound structures in the simulations.
For bound groups the actual fraction is indistinguishable from the 
recovered fraction, which simply illustrates that all bound groups 
are correctly recovered by our clustering scheme. The dashed 
lines, i.e., the actual fractions, show that outer halo (beyond 120 kpc) 
is predominantly bound whereas the inner halo (less than 60 kpc) 
is predominantly unbound. Next we look at the recovered fractions.
The $f^{\rm recovered}_{\rm unbound}(r)$ is quite low in the inner
regions and then rises slowly and peaks at around 100
kpc. This rise coincides with a fall of fraction $f^{\rm
  recovered}_{\rm bound}(r)$. 
This is due to the following two facts 
a) low $\epsilon$ (circularity of orbit) are more likely to be 
unbound than circular events since they pass close to the center,  
b) low $\epsilon$ events are more likely to be found at the apocenter 
and their apocenter is further away than that of an high $\epsilon$ event
of similar energy.
The black line representing $f^{\rm recovered}_{\rm bound+unbound}(r)$ shows 
that the outer halo (beyond 60 kpc) is highly structured whereas 
the inner halo (less than 40 kpc) which contains about 50\% of the
material (as shown by the red line) has very little material 
which can be recovered
as structures. This implies that 
in the inner regions strong phase mixing greatly 
limits the amount of material that can be recovered as structures in 3-d
configuration space. The analysis also suggests that 
for a given survey the fraction of material in groups  
will depend sensitively upon the contribution of the inner 
halo to the survey, which in turn is determined by the geometry 
and the depth of the survey.  

A few other issues which can affect the fraction of material in groups  
are as follows. We have here considered only the 3-d configuration
space, additional information in the form of velocities 
should help detect more groups and increase the fraction of material in 
groups. 
Also, the choice of photometric selection function can decrease 
or increase the fraction of material in groups depending upon the 
contribution of the smooth component to the sample of stars in the
survey (see \sec{events} for further details).
Issues related to our choice of the clustering scheme which
can affect the fraction of material in groups are discussed in \sec{SDSS}.

 \begin{figure}
   \centering \includegraphics[width=0.49\textwidth]{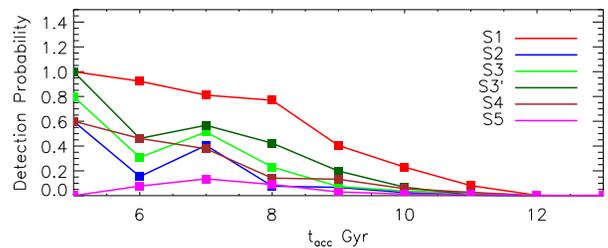}
 \caption{ Detection probability of accretion events as a function of
   time for different data sets.  
The detection probability was computed by binning the accretion events 
along time in bins of $1 \Gyr$ and subsequently computing the  
ratio of the number of unique accretion events detected by the 
group finder to the total number of events in a bin. 
Data from all 11 $\Lambda$CDM halos was used.
 \label{fig:f7}}
 \end{figure}

\begin{figure}
  \centering \includegraphics[width=0.49\textwidth]{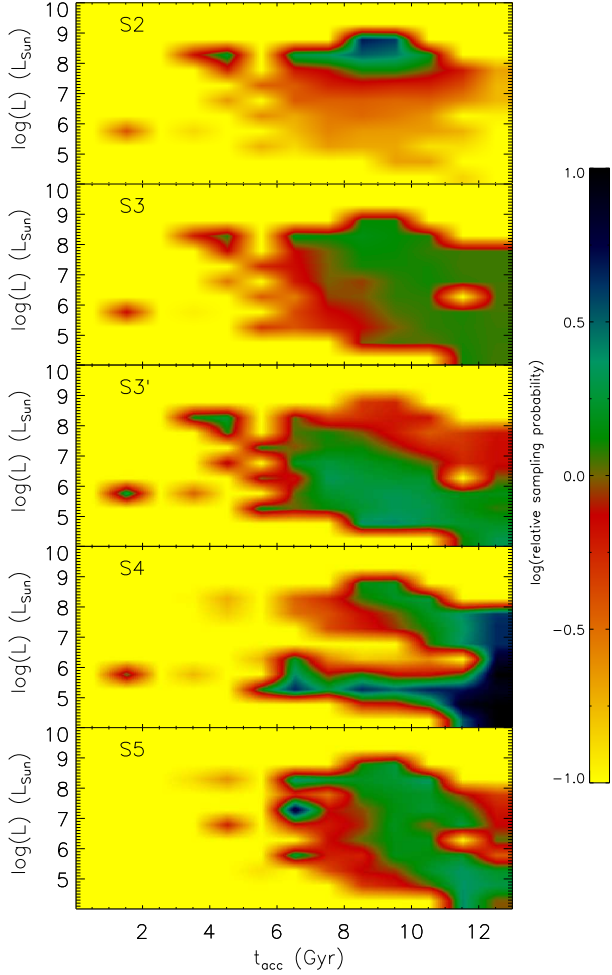}
\caption{ Sampling probability of accretion events in the $(L,t_{\rm acc})$  
parameter space for different photometric surveys relative to data set
S1. Different data sets are labeled on the plot. 
\label{fig:f8}}
\end{figure}

\begin{figure}
  \centering \includegraphics[width=0.49\textwidth]{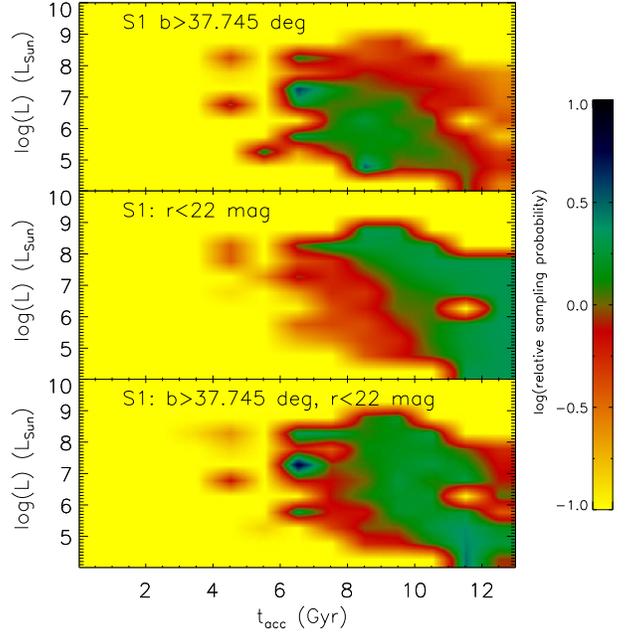}
\caption{ {Sampling probability of accretion events in the $(L,t_{\rm acc})$  
parameter space for different photometric surveys relative to data set
S1. Different data sets are labeled on the plot and are subsamples 
of data set S1. The plots shows the effect of magnitude limits and 
latitude limits on the sampling probability.
The top plot is for 
latitude restricted to $b>37.74$, the middle panel is for
SDSS $r$ band magnitude limited to $r<22$ and the bottom panel is for both 
$r<22$ and $b>37.74 \degree$. 
\label{fig:f9}}}
\end{figure}

\section{Results II: surveys in a realistic Universe}\label{sec:results2}

In this section we apply the group-finder to our more realistic
synthetic surveys of stellar halos (data sets S2--S5), compare their
sensitivity to different accretion events and assess their ability to
distinguish accretion histories.  We apply the group-finder with
$k=30$ and selecting $S_{\rm Th}$ according to \equ{GS} to
account for the difference in the sample sizes, to all seventeen
stellar halo models within each data set.

\subsection{The effect of stellar populations}\label{sec:stellarpop}
\fig{f7} summarizes our results by plotting the number fraction of
accretion events recovered as a function of accretion time, where each
symbol represents a 1 Gyr interval. As expected, our idealized
survey S1 recovers the largest fraction of events at all times, and
the SDSS MSTO survey (S5), which covers less than 1\% of the volume of
any of the other surveys, recovers the fewest.  However, there is no
simple answer to the question of which survey is most sensitive.  Our
synthetic data sets represent surveys of different stellar tracers
(i.e. M-giant, MSTO or RR Lyrae stars) selected with different
observing strategies.  As a result, these data sets explore the space
around the Galaxy with a variety in the numbers of stars, depths and
accuracies in distance estimates, as outlined in \sec{sim2sur} and
summarized in \tab{tb1}.  In addition to these differences in spatial
exploration, systematic differences in the stellar populations of
objects of different luminosities and accretion times \citep[e.g. as
reflected in the stellar-mass/metallicity relation for Local Group
dwarfs, see][]{1974MNRAS.169..229L}
means that the choice of tracers affects the relative
number of stars contributing to each data set from different accretion
events.  

\fig{f8} illustrates this effect by plotting the sampling probability  
of accretion events in the $L$-$t_{\rm acc}$ plane relative to data
set S1-- the sampling probability being computed as the probability 
distribution of stars appearing in a survey in the 
$L$-$t_{\rm acc}$ plane (the $L$ 
and $t_{\rm acc}$ of a star being that of its parent satellite). 
The top panel, for data set S2, clearly shows the expected bias of
M-giants towards tracing the highest metallicity and hence highest
luminosity events.  In addition, M-giants are intermediate age stars,
which means that such surveys do not contain stars from ancient
accretion events.

The second two panels show the sampling probability 
for our deep MSTO surveys S3 and S3',
which differ only in the color range from which stars are selected.
Both surveys contain stars from lower luminosity objects and earlier
accretion times than the S2 surveys. A comparison of the two shows
that as the red edge of the color limit increases, the sampling probability
increases for old and high luminosity events (upper right hand region
in the plots) and decreasing the red edge of the color limit has the
opposite effect, i.e., sampling probability increases for recent and low
luminosity events (lower left hand region in the plots).  This is
because a) the blue edge of the MSTO stars in an isochrone (knee
shaped feature in the color magnitude diagram) shifts red-wards with
the increase in age and metallicity of the stars and b) the high
luminosity events are also metal rich.  Increasing the photometric
errors has an effect similar to increasing the color
range. \footnote{Increasing the photometric errors, also results in an
  increase in contamination from red dwarfs, which have low magnitude
  and hence high photometric errors.}
This is the reason why the sampling probability for high luminosity events 
is slightly higher for S3 as compared to S1 although both surveys 
have the same color limits.

The fourth panel, for the RR Lyrae survey (data set S4), shows the
strongest bias towards old and low-luminosity events as RR Lyraes
are old, low-metallicity stars.

The bottom panel, for the MSTO samples with SDSS sky-coverage and
magnitude limits (i.e. data set S5), indicates that stars from recent,
low-luminosity events and ancient, high-luminosity events are poorly
represented in this survey compared to the S1 sample. Recall that S1
surveyed all-sky to greater depth (r=24.5) than S5, with the same
color cut, but assuming perfect knowledge of distances.  \fig{f9}
shows the sampling probability for surveys like S1 but with an
SDSS-like latitude (top panel) and magnitude (middle panel) cut.
Combining the two cuts (bottom) panel reproduces the bulk of the
features in the lower panel of \fig{f8} which demonstrates that
accretion events are ``missing'' from the SDSS sample simply because
of the small sample volume rather than errors in distance estimates.

Overall, we conclude that the variety of observing strategies and tracers
used in current and future surveys
can lead to a range in the contrast that a given structure
may have with respect to the smooth background. 
This suggests that different surveys will be sensitive to
different types of accretion events.

\subsection{Events recovered as groups} \label{sec:events}

The last five rows in \fig{f4} show the fraction of events recovered as
groups from our S2, S3, S3', S4 and S5 surveys of the $\Lambda$CDM 
stellar halo models, which can be broadly understood given the survey
characteristics outlined in \tab{tb1} and \sec{stellarpop}. 

For example, relative to the other surveys, the S2 (M-giant) data sets
contain a low number of stars, that explore only out to $\sim$100 kpc 
from the center of the Milky Way with modestly accurate distance
estimates and a bias towards intermediate age, high-metallicity
populations.  These effects combined, explain the absence of groups in
the S2 panels of \fig{f4} detected from events that were: old (M-giant
stars are intermediate age); low luminosity (M-giant stars are
relatively high metallicity); or on eccentric orbits (with apocenters
beyond 100 kpc where debris spends most of its time).  However, despite
the relatively low number of stars overall, these surveys are
particularly well-suited to finding signatures of high luminosity,
relatively recent accretion events: these stand out in M-giants at
high surface brightness (because of the high-luminosity and recent
accretion of the progenitor) and 
better 
contrast to the background (since ancient events which form a smooth 
background are not represented).

Data set S3 (MSTO stars) explores a similar volume to the S2 surveys, but
contains a far greater
number of stars, with less accurate distance
estimates and a wider variety in the ages and metallicities of stars.
Overall \fig{f4} demonstrates that these MSTO surveys can recover
additional groups from lower luminosity and more ancient accretion
events (because the survey contains lower metallicity and older
stars), as well as those on more eccentric orbits (because the large
distance errors mean that a significant fraction of stars scatter into
the sample from beyond 100 kpc).

Data set S3' is constructed the same way as S3, but with a finer color
cut. This lowers the total number of stars, increases the accuracy of
the distance estimates but decreases the contribution from older and
higher metallicity stars.  Since the last property decreases the
contribution of the smooth background, the S3' surveys do even better
than the S3 surveys at recovering groups corresponding to low
luminosity events.  \footnote{Note that we found that decreasing the
  photometric errors --- as possible with co-added LSST data ---
  emphasizes the trends found with making a finer color cut as there
  is less scatter in the colors. However, it is hard to assess the
  importance of this consideration given the simplicity of the stellar
  populations in our models.}

Data set S4 --- the RR Lyrae surveys --- contain only a comparable
number of stars to the M-giants surveys, but explore to far greater
depth and with much more accurate distances than S2, S3 or S3', and
represent old, low-metallicity populations. \fig{f8} shows that stars
from intermediate-luminosity satellites are missing from the survey,
and \fig{f4} confirms that these events are not recovered as groups.
However, S4 has a similar success as S3 and S3' in filling out other
regions of accretion history space.

Finally, data set S5 --- the SDSS MSTO surveys --- contains the same
stellar populations as S3, but explores a much smaller volume because
of the brighter magnitude limit. A comparison of the bottom panels of
\fig{f4} with and \fig{f8} demonstrates how this shallower magnitude
limit effectively eliminates sensitivity to many recent events which
do not contribute significantly to the inner halo. Similarly, the
bottom right-hand panel of \fig{f4} shows that the group finder
recovers no events on eccentric orbits that are likely to have debris
beyond the edge of the SDSS survey.

\subsection{Mapping accretion history with combined surveys}

Recall that in our idealized surveys we found that number of groups
(\tab{tb2}), amount of material in substructures (\tab{tb3}) and the
probability distributions of group masses (\fig{f5}) and radial
distance (\fig{f6}) can be related to the number, luminosity function
and orbit distribution of recent accretion events.  If we now account
for the sensitivity of the different surveys to accretion events of
different properties outlined in the previous two sections we can
examine how such surveys could combine to present a picture of recent
accretion history. Note that, because of the low number of groups
recovered, we have not included the results for the distribution of
group properties in data set S5 in \fig{f5} and \fig{f6}.

For example, our S2 M-giant surveys are not expected to be sensitive
to low luminosity events or those on very radial orbits. This
insensitivity accounts for why \tab{tb2} shows that almost no groups
were recovered from these surveys of our mock-halos built entirely
from radial orbit events, and why there is not a striking difference
in the number of groups recovered from our low-luminosity model and
our standard $\Lambda$CDM models.  However, the S2 panel in \fig{f5}
clearly shows that our high-luminosity model halo gave rise to a much
larger fraction in massive groups than our $\Lambda$CDM models.
Hence, we expect that 2MASS can already provide a reasonable census of
recent accretion events from the upper end of the
luminosity-function.

Previous results for our S3 and S3' data sets suggest that information
about the lower end of the luminosity function of recent accretion
events might be filled in by these future MSTO surveys.  Indeed, \tab{tb2}
indicate that these surveys detect a 
larger number of groups in our low-luminosity halo, which
contained a larger number of low-luminosity accretors, when compared
to our standard models.  This indicates that MSTO surveys should
recover debris from accreted objects --- like the lower luminosity
classical dwarf spheroidal and ultra-faint galaxies --- that would be
missed by 2MASS.

Since they explore to far greater depth, future RR Lyrae surveys (S4)
will have unique power to probe the outer halo, beyond 100 kpc and
assess the fraction of debris that lies at these large radii. This
power is demonstrated by the very clear difference in the S4 panel of
\fig{f6} between the radial distribution of groups found in the RR
Lyrae survey of our radial orbit halo (magenta line) compared to the
$\Lambda$CDM cases (thin black lines). This difference implies that RR
Lyrae surveys will flesh out our picture of the orbit distribution of
accretors onto our Galaxy. Additionally they are also very efficient
at detecting the low luminosity accretion events. 

\subsection{Limitations}
The synthetic surveys that we analyze do not include other 
galactic components, e.g., the disk and the bulge. However,
this will have little impact on the analysis we present here because  
a) the volume of the stellar halo that we explore (100 kpc and beyond) 
is much larger as compared to the volume of the disc. b) most stellar
halo structures lie away from the disc and bulge regions.  
Note, even in the present analysis we have neglected groups that lie
within  5 kpc of sun or the galactic center (\sec{group_finder}). 
Another thing that we have neglected is the extinction which will 
make the detection of halo stars in the plane of the disc very difficult.
In realistic surveys one might have to filter out the low lying 
latitudes  as was done for real 2MASS data in Paper-I. 
However an analysis of S2 data after removing the low latitude regions 
revealed that this has little impact on the number of reported 
groups or the ability to discern different accretion history. 
In general a few structures will be missed while a few others 
will be split and discovered as two groups.  

For the MSTO surveys we have neglected the possibility of 
contamination by quasars (QSO). 
They are expected to contaminate towards the blue color limits.  
In the color range $0.2<g-r<0.4$ and $r<21.5$ 
\citet{2008ApJ...673..864J}  estimate the contamination 
by QSO's to be 10\%. Since QSO's would be distributed 
isotropically and in a smooth fashion they will not give rise to 
any false structures but due to their contribution to the background 
they might decrease the significance of some of the faint structures 
making them difficult to identify.

A final limitation of our analysis is the fact that recent accretion 
events with $L<10^5 \Lsun$ are not included in the
 stellar halo simulations that we use here. This is
because  the star formation recipes in the simulations were fine 
tuned to reproduce the observed distribution of satellites known at
the time of the publication of \citet{2005ApJ...635..931B}, 
and the ultra faint galaxies were discovered later on.  
Our analysis shows that the MSTO and RR Lyrae surveys 
are sensitive to low luminosity events hence we anticipate such
surveys to detect more structures than that reported by us.

\begin{figure}
  \centering \includegraphics[width=0.49\textwidth]{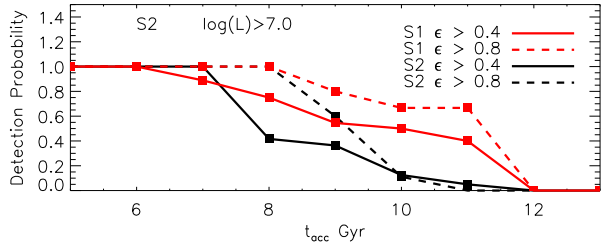}
\caption{ Detection probability of accretion events restricted 
to $L >10^7 \Lsun$ and high circularity ($\epsilon>0.4$ or $\epsilon>0.8$). 
Shown are the probabilities for data set S1 and S2. 
It can be seen that circular 
events ($\epsilon >0.8$ as compared to $\epsilon >0.4$) can be detected much further back in time. Comparison of the 
curves for S2 with that of S1 illustrates that the 2MASS survey is 
almost complete for high luminosity and high circularity events.  
\label{fig:f10}}
\end{figure}

\section{Applications to the real Universe}\label{sec:results3}

\subsection{The 2MASS M-giant sample}

In Paper I we identified structures in the 2MASS M-giant sample and
found sixteen groups, seven of which corresponded to known bound
structures, one was probably a part of the disk and two others were
due to masks employed in the data. Excluding these, we found six
unbound groups.

The 2MASS results in \tab{tb2} and \tab{tb3} and in bold dashed
lines in \fig{f5} and \fig{f6} provide a detailed comparison between
the properties of these observed groups and those recovered from the
synthetic M-giant surveys (data set S2) of our eleven $\Lambda$CDM
models (thin black lines).  Overall, we find broad consistency between
the observations and models.  In particular, the total number of
groups (six) sits close to the middle of the range found in the
models (8.4$\pm$ 8).

Note the results reported for data set S2 are without any selection
by latitude. We also repeated the analysis by applying the same
latitude limits used to generate our 2MASS M-giant sample but did not find
a significant change in the mean number of recovered groups for
$\Lambda$CDM halos, or in the figures.  However, the fraction of material
in groups was found to be around 30\%, which is about 3 times higher than that 
reported in \tab{tb3} (unbound groups). This is because much of the stellar halo mass is 
concentrated towards the center and the plane of the galaxy. 
This mass near the center of the
galaxy does not contain any substructures and its exclusion results in
an increase in the mass fraction of recovered groups. 
On the other hand in the real 2MASS M-giant data we find the mass
fraction in groups to be 4.5\%, which is quite low. The reason for this 
is that in the real 2MASS data nearly 75\% of the stars are due to the 
LMC which is a bound satellite system and this is quite un-typical 
for a $\Lambda$CDM halo. 
For comparison in our simulated halos the maximum fraction of 
mass in a single bound satellite was 20\%. Excluding LMC, increases  
the fraction of material in groups to 18\%, which is more typical of what
we see in simulated halos.  

We can also use our synthetic surveys to ask whether the 2MASS groups
are likely to correspond to real accretion events and, if so, what
type of events.  Table 1 shows that the typical purity of groups
recovered from the simulated halos in data set S2 was high, which
supports the interpretation of the groups in 2MASS as real physical
associations. Moreover, the results from \sec{results2} suggest that
these groups correspond to recent ($< 10 \Gyr$ ago), high-luminosity
accretion ($> 5 \times 10^6 \Lsun$) events on orbits whose
apocenters lie within 100 kpc.  

Finally, we can test what the 2MASS groups can contribute to our
picture of accretion history by asking what fraction of accreting
objects we expect to detect as groups in this data set.  \fig{f7}
suggest the groups represent 60\% of objects accreted within the last
6 Gyr, falling to as little as 10\% of those accreted 7.5-8.5 Gyr
ago and 0\% of those with accreted earlier than 11.5 Gyr ago.
While these numbers clearly portray the weakness of the M-giants as
tracers of accretion history, they do not capture their strength.
\fig{f10} repeats \fig{f7}, but this time restricting attention to
events that we know contain M-giants within the apparent magnitude
range of the 2MASS survey --- those that have high-luminosity ($>10^7
L_\odot$) and are on mildly-eccentric orbits ($\epsilon > 0.8$). 100\%
of these events are recovered that were accreted within the last 8
Gyr, and more than 50\% with accretion times less than 10 Gyr ago.
We conclude that the structures in 2MASS give us a fairly complete
census of recent, massive accretion events along mildly-eccentric
orbits.

\subsection{SDSS MSTO sample} \label{sec:SDSS}
While a full application of our group finder to the SDSS MSTO sample
is beyond the scope of the current work, analysis of our S5 data sets
suggest we might expect to find 0-4 groups, corresponding to
relatively high-luminosity, intermediate-age accretion events on
mildly eccentric orbits. The S2 and S5 panels of \fig{f4} imply that
the majority (though not all) of these events would also be recovered
from the 2MASS M-giant sample. Indeed, 4 groups that are plausibly
associated with the ancestral siblings of the classical dwarf
spheroidal satellites are apparent through visual inspection of SDSS
MSTO sample \citep[the tails from disruption of Sagittarius, the
Monocerous ring and the Virgo overdensity, and the Orphan Stream, see
e.g.][]{2006ApJ...642L.137B}, and two of these are also seen in the
M-giants.  This again indicates broad consistency between structures
seen in the real stellar halo and those in our models.  \citep[While
many more streams from globular clusters and lower-luminosity dwarfs
have been found in SDSS using matched-filtering techniques, e.g.,][
these objects are missing from the stellar halo models, so we compare
only the number of higher-luminosity streams]{grillmair09}

From \tab{tb3} it can be seen that, for data set S5,  
the mean fraction of material in groups 
 is  4\% which increases to 6\% when bound groups are 
also included. These results are
 consistent with what we see for S3 survey, if we take into account 
the shallow depth of the S5 survey due to which the outer halo, which
is also more structured (see \fig{fraction}), is missed . Similarly, most bound
structures are in the outer parts of the halos (see \fig{fraction}), 
hence unlike other surveys, including them does not increase the
fraction by much for S5 survey. 
However, when compared to the results 
of \citet{2008ApJ...680..295B}, where they find the fractional rms 
deviation from  a smooth analytic model, $\sigma/{\rm total}$, 
to be around 40\% for SDSS MSTO stars, our mass fractions are
apparently low. 
It should be noted that \citet{2008ApJ...680..295B} had 
also analyzed the same set of halos as used by us and reported 
good agreement with the SDSS data. Hence the cause of the mismatch 
is due to the methodologies being different, which we explore 
below. Firstly,  we think that $\sigma/{\rm total}$ 
cannot be directly interpreted as amount of mass in structure, e.g.,   
considering equally populated bins, if 10\% of the bins differ in 
mass by order of the mass in the bins one gets $\sigma/{\rm total}=0.33$.
Secondly, an analytical model as adopted by them 
might not necessarily be a good description of the smooth component 
of the halo. This misfit will contribute to  $\sigma/{\rm
  total}$ but will not give rise to any structure in our scheme.
It is also important to note that in our clustering scheme only 
those structures are detected which give rise to peak in the 
density distribution.
Thirdly, the group 
finder truncates a structure when its isodensity 
contour hits that of another structure.  
Hence the envelope region 
around a structure can contribute to   $\sigma/{\rm total}$ 
but is ignored by us. Note, by following points in the envelope
regions along density gradients  
one can associate them with the groups but such a classification 
is not free from ambiguity and in general decreases the purity of the groups. 
Finally, in our analysis we only consider significant groups 
which stand out above the Poisson noise. However, there might be low 
significance fluctuations which we have deliberately ignored 
and these will invariably contribute
to $\sigma/{\rm total}$. However, low significance
fluctuations unless they are massive, which is rare, wont 
dominate the mass fraction in structures. 
To conclude we think that the methodologies 
being very different it is very difficult to 
interpret the results of one scheme in terms of the other.

% --------------------------------------------------------------------------------
\section{Summary}\label{sec:summary}
In this paper, we have explored the power of a group finding algorithm
to recover structures from photometric surveys of the stellar halo and
interpret their properties in terms of Galactic accretion history.

We first applied our group finder to idealized synthetic stellar
surveys, which were generated from our $\Lambda$CDM models without
accounting for observational errors.  We find a simple dependence for
the probability of detecting debris as a group on the parameters of
its progenitor accretion event: the probability is highest for recent
(small $t_{\rm acc}$) and high luminosity (large $L$) events accreted
along circular orbits (large $\epsilon$). The strongest dependence is
on $t_{\rm acc}$.  The properties of recovered groups --- the number
of and fraction of material in groups, along with distribution of the
stellar mass and radial distance of the groups --- can in principle
place constraints on the {\it recent} accretion history of a
halo. Ancient accretion events ($> 10\Gyr$ ago) are not recovered as
groups even in the absence of observational limitations because they
are too phase-mixed to appear as distinct structures.

We then applied our group finder to synthetic surveys that contained
more realistic observational errors. Our results emphasize that the
capability of a photometric survey to discover structures depends upon
its sample size, the distance uncertainty, the depth and the relative
sampling probability of different stellar populations.  The
broadest constraints on accretion history will come by combining the
results of current and future surveys.  For example, M-giants selected
from 2MASS are intermediate age, high metallicity stars and hence are
good tracers of relatively recent, high-luminosity accretion events with little
contamination from older events.  An LSST MSTO survey would contain a
range of stellar populations whose properties depend on the severity
of the color cut made to select the sample.  Limiting the sample to
the very bluest MSTO stars increases the dominance of low-metallicity
stars in the sample, and hence the sensitivity to low-luminosity
accretion events.  RR Lyraes selected from LSST are bright
enough to probe beyond 100 kpc where the apocenters of the more
eccentric orbits lie, and hence will find a more fair sampling of the orbital
properties of accretion events.

Finally, a quantitative comparison of the results of applying the
group-finder to the real 2MASS M-giant sample with those from the mock
(synthetic) 2MASS M-giant surveys shows the number and properties of
substructures in 2MASS M-giant survey to be roughly in agreement with
simulated $\Lambda$CDM stellar halos. These groups most likely
correspond to satellites accreted more recently than $10 \Gyr$, with
luminosity higher than $5 \times 10^6 \Lsun$ and preferably on 
orbits of low eccentricity.

Overall we conclude that current and near-future photometric surveys
are poised to provide a complete census of our Galaxy's recent
accretion history. The current results from 2MASS alone map the
highest-luminosity recent events, future deep MSTO surveys will fill
in the lower end of the luminosity function and RR Lyrae surveys will
find debris structures that may be currently missing because the
progenitor satellites were on highly radial orbits.
Reconstructing more ancient accretion will require additional
dimensions of data, such as  velocity (proper
motions and radial velocities of stars) and chemical abundance information.

\section*{Acknowledgments}
This project was supported by the {\it SIM Lite} key project {\it Taking
  Measure of the Milky Way} under NASA/JPL contract 1228235. SRM and  RRM
appreciate additional support from NSF grants AST-0307851 and AST-0807945. We
also thank Brant Robertson and Andreea Font, whose work on chemical enrichment
made it possible to calculate the age-metallicity relation for star particles
in simulations.

\providecommand{\newblock}{} 
\bibliographystyle{apj} 
\bibliography{/home/sharma/texmf/mybib}
%\bibliography{ms}

\appendix
\section{Effect of sub-sampling} \label{sec:subsampling}
In \sec{sur} we had mentioned that for 
computational ease  we had sub-sampled the data sets S1 and S3 
by a factor of 0.25. This sub-sampling resulted in a sample size of about
$10^7$ while the actual sample size would have been $4 \times 10^7$.  
To check if sub-sampling has any effect on the results of our group 
finding analysis, we ran the group  finder on the eleven
$\Lambda$CDM halos with ten and 100 times lower number of stars, 
i.e., sample sizes of $10^6$ and $10^5$ respectively. The parameter 
$S_{Th}$ was set according to \equ{GS}. 
For data set S1 lowering the resolution by ten and 100 times 
resulted in a reduction in the number of detected groups by 14\% and 64\% 
respectively while the purity was found to be remain approximately constant. 
This suggests that increasing the sample size from $10^5$ to
$10^6$ results 
in significant improvement in clustering performance but beyond $10^6$ 
the performance tends to saturate.
This saturation of clustering performance is partly due to the 
fact that in the simulations there is a finite 
lower limit on the luminosity of the satellites that are simulated.   
Similar trends were also 
seen for the data set S3-- lowering the resolution by 10 times
resulted in a reduction in number of groups by 10\%.
Hence we conclude that choosing $f_{\rm sample}=1$ instead
of 0.25 as done by us should only give a moderate improvement in
clustering performance.

\end{document}